# Enhanced Wall Boundary Modeling for Turbulent Flows Using the Lattice Boltzmann Method with Adaptive Cartesian Grids

**Jorge Ponsin\*, Carlos Lozano**

*Theoretical and Computational Aerodynamics*

*National Institute of Aerospace Technology (INTA)*

*Torrejón de Ardoz 28850, Madrid, Spain*

*\*Corresponding author: ponsinj@inta.es*

## Summary

We propose an enhanced wall-boundary treatment for the lattice Boltzmann method (LBM), designed for high-Reynolds-number turbulent flows on adaptively refined Cartesian grids. The method improves the slip-velocity bounce-back scheme by coupling it with a near-wall turbulence model based on an analytical wall function. The Spalart-Allmaras (negative) turbulence model is solved using a second-order finite-difference scheme and integrated within the LBM framework to statistically represent the Reynolds-Averaged Navier-Stokes (RANS) equations (LBM-RANS). The approach is validated on two benchmark configurations: the National Advisory Committee for Aeronautics (NACA) 0012 airfoil and the McDonnell Douglas (MD)-30P30N multi-element high-lift configuration. LBM-RANS results show good agreement with conventional finite-volume RANS solutions and experimental data for key aerodynamic quantities, including pressure and skin-friction distributions, as well as turbulent boundary layer velocity profiles and eddy-viscosity fields. The method delivers smooth and accurate predictions of skin friction, which are often challenging for immersed-boundary approaches on Cartesian grids. The auxiliary geometric data required for enforcing the turbulent boundary condition are minimal, making the method potentially well-suited for graphics processing unit (GPU)-based implementations. Moreover, no ad-hoc near-wall treatments are needed, as the boundary condition is applied naturally via the link-wise bounce-back scheme. These results illustrate that the proposed LBM-RANS framework can robustly and accurately simulate high-Reynolds-number turbulent two-dimensional (2D) flows over complex aerodynamic geometries under equilibrium or near-equilibrium conditions.

## 1. Introduction

### 1.1 Background and context

Over the past two decades, the Lattice Boltzmann Method (LBM) has emerged as a promising alternative to traditional computational fluid dynamics (CFD) techniques for simulating aerodynamic flows, particularly at high Reynolds numbers using scale-resolving turbulence models. Its appeal lies in the method's inherently low numerical dissipation, combined with the efficiency of the stream-collide algorithm, making it well-suited for capturing unsteady, multiscale flow phenomena. These advantages have been demonstrated in several industrially relevant applications [1] [2] [3] [4] [5] [6].

Standard LBM implementations are based on isotropic Cartesian lattices, which enable the efficient stream-collide procedure. However, flows involving complex geometries and disparate length scales require multilevel adaptive Cartesian grids to maintain computational feasibility, as uniform grids would lead to prohibitive computational costs. For high-Reynolds-number turbulent flows, near-wall modeling is also essential, since fully resolving the near-wall region would demand extremely fine grids. Accurate wall-treatment strategies capable of incorporating turbulence wall models within the Cartesian grid framework are therefore critical for extending LBM to reliable high-Reynolds-number simulations.

Despite significant research efforts, the development of robust wall-boundary conditions for LBM remains an open challenge [7]. The literature offers no clear consensus on the most effective turbulent wall treatment for the LBM framework. Most existing approaches employ ad hoc mesoscopic prescriptions for distribution functions at boundary nodes, complicating the formulation of consistent macroscopic boundary conditions. Theoretical analyses linking mesoscopic boundary schemes to their macroscopic counterparts are scarce and have largely been limited to laminar flows and simple geometries [8] [9] [10]. This challenge becomes even more pronounced when combining LBM with turbulence wall models and adaptive Cartesian grids. The recent proliferation of wall-boundary treatments for high-Reynolds-number turbulent LBM simulations underscores both the complexity of the problem and the current lack of a universally accepted solution.

### 1.2 Wall-boundary treatments for turbulent flows in LBM

Wall-boundary treatments for LBM with turbulent wall modeling can be broadly classified into three categories, based on the type of the mesoscopic boundary condition employed: wet-node, bounce-back, and volumetric approaches.

#### 1.2.1 Wet-node approach

Wet-node approaches replace all particle distribution functions at boundary nodes, whether located in fluid or solid regions, with reconstructed equilibrium and non-equilibrium distributions [7]. The equilibrium part is typically assembled using the local density and tangential velocity at the boundary node, with the latter obtained from a wall-function formulation and the density extrapolated from the fluid domain. Interpolation or extrapolation is generally required to handle Cartesian grid topologies. The key difference among wet-node approaches lies in the reconstruction of the non-equilibrium distributions. Two main strategies have been developed for turbulent flows:

1. Non-equilibrium distribution extrapolation. Originally introduced by Filipova et al., [11] the method relies on a link-wise bounce-back formulation to reconstruct the non-equilibrium distribution functions at solid nodes. Subsequently, Guo [12] used a wet (fluid) node to extrapolate the non-equilibrium distributions, an idea that was later extended by Haussmann et al. [13] [14] and Maeyama et al. [15] [16] [17] in the context of scale-resolving simulations. It is worth noting that although some of these approaches employ a link-wise bounce-back procedure for reconstructing the non-equilibrium distributions, they are not classified within the bounce-back family (as described later) since the bounce-back mechanism is used solely for reconstructing the non-equilibrium components rather than for enforcing boundary conditions.

2. Regularized reconstruction, introduced by Malaspinas et al. [18] for flat, grid-aligned walls and later extended to curved geometries [19] [20]. The second strategy has been successfully applied within the framework of LBM-RANS simulations. In this regularization-based approach, the non-equilibrium populations are approximated using the first-order term of the Chapman-Enskog expansion, reconstructed from local density and velocity gradients at the boundary node. Accurate estimation of the wall-normal velocity gradient is particularly critical for achieving reliable predictions, as demonstrated in [21]. Early applications to high-Reynolds-number flows on Cartesian grids were reported by Wilhelm et al. [19] and Cai et al. [20], who achieved reasonable predictions of pressure and skin-friction distributions but still observed spurious oscillations in surface quantities. Degrigny et al. [7] [22] proposed an improved wall treatment incorporating multiple refinements, such as discarding nodes too close to the wall, applying third-order finite-difference schemes for wall-normal gradients, and introducing special handling of discontinuities. These improvements, implemented in the ProLB solver, significantly enhanced solution smoothness and the accuracy of surface pressure and friction predictions in LBM-RANS simulations. More recently, Husson et al. [23] conducted a comprehensive investigation of the impact of wall-model data interpolation within a regularized treatment in the LBM-RANS framework, emphasizing the importance of selecting appropriate interpolation strategies when modeling turbulent flows in the presence of immersed walls. Recent applications of these turbulent wall treatments in ProLB include the work of Mozaffari et al. [24] , who employed LBM combined with a hybrid RANS-Large Eddy Simulation (LES) approach and a wall-function model accounting for pressure-gradient effects, as well as the studies by Husson et al. [25] [26] within the context of LBM coupled with the Zonal Detached Eddy Simulation (ZDES) turbulence model.

**1.2.2 Bounce-back approach**

The second category comprises bounce-back approaches. In the classical bounce-back method [27] [28], only incoming populations are replaced by their counterparts from opposite lattice directions after reflection at the wall. Owing to its simplicity and robustness, this approach is widely used in laminar-flow simulations [27], but its application to high-Reynolds-number turbulent flows with wall models has been relatively limited. Notable examples include, on the one hand, the works of Pasquali et al. [29] and Gehrke et al. [30], who applied the method to turbulent channel-flow prediction, and, on the other hand, the studies of Ishida [31] [32], Nishimura et al. [33] and Liang et al. [34], who employed a slip-velocity bounce-back scheme for aeroacoustic noise prediction of the turbulent flow around the MD-30P30N high-lift airfoil. All the aforementioned studies were conducted within the framework of wall-modeled large-eddy simulation (WMLES). By contrast, considerably fewer investigations have addressed LBM-RANS formulations. Among these, the work of Lyu et al. [35] is particularly noteworthy.

The approaches proposed by Pasquali [29] and Gehrke [30] are closely tied to the cumulant collision model, which limits their portability to other LBM solvers. Pasquali introduced the concept of a partial-slip velocity scheme combined with an implicit WMLES (iWMLES) formulation for turbulent channel-flow prediction. More recently, Gehrke applied the inverse momentum-exchange method (iMEM) boundary scheme [36] to channel-flow simulations, achieving improved predictive performance compared with earlier implementations. Nevertheless, he emphasizes that the geometric generalization of this wall treatment remains an open challenge.

Ishida [31] employed an interpolated bounce-back scheme coupled with a wall function to compute the slip velocity used in the wall-momentum term of the bounce-back formulation. However, no details are provided regarding the specific wall-function model adopted or the manner in which the coupling is implemented. Using an iWMLES turbulence approach, he simulated the flow around the MD-30P30N airfoil in both two and three-dimensional configurations. The 2D results were physically unrealistic, while the 3D simulations focused primarily on sound spectra; consequently, no assessment of the wall-model accuracy for predicting the near-wall turbulent flow was reported. Nishimura et al. [33] presented a formulation that is conceptually similar to the method proposed here. They performed 3D iWMLES simulations of the MD-30P30N at a Reynolds number of $1.7 \times 10^6$ and obtained good agreement with experimental data for the time averaged pressure distribution. However, they did not provide a detailed validation of the turbulent boundary-layer characteristics, such as skin friction coefficient distribution or velocity profiles, focusing instead on noise generation in the slat-cove region. As a result, the reliability of the wall treatment for turbulence modeling remains difficult to evaluate. A more recent study building on Nishimura's formulation was carried out by Liang et al. [34], who also applied the method to aero-acoustic simulations of the MD-30P30N airfoil. They used the Smagorinsky subgrid-scale (SGS) model and presented results for both 2D and 3D configurations. The 2D pressure distributions were poorly predicted, as the computed flow was physically unrealistic. The 3D results also exhibited some discrepancies in the predicted pressure distribution. As in the works of Ishida and Nishimura, the analysis primarily focused on aeroacoustic noise, and no further evaluation of the wall-model accuracy was provided.

Lyu et al. [35], working within the framework of LBM-RANS simulations, introduce a new hybrid local grid refinement method in the near-wall region that combines different resolution levels around the wall to increase simulation efficiency. Their approach employs the $k-\varepsilon$ RNG eddy-viscosity model together with the wall function proposed by Cai et al. [20]. Although the authors describe the extrapolation procedures used to reconstruct flow variables at boundary nodes, they provide no details regarding the coupling between the wall model and the bounce-back scheme. Instead, the study concentrates primarily on the interpolation treatment applied at the transition between different resolution levels around the wall. The authors present results for the surface-pressure distribution over a NACA0012 airfoil at a Reynolds number of $6\times10^6$ and an angle of attack of 10º. However, it is difficult to assess the accuracy of the proposed wall-model coupling, as no detailed validation of turbulent quantities (such as skin-friction distributions, boundary-layer velocity profiles, or turbulence levels) is provided.

**1.2.3 Alternative boundary treatment approaches**

In the third category, we briefly describe other specialized approaches that are not included in the categories mentioned above. The volumetric approach was initially proposed by Chen et al. [37] and later extended to turbulence RANS modeling with wall functions in the pioneering work of Teixeira [38]. This approach uses the native surface tessellation and auxiliary control volumes to compute particle distribution dynamics between wall and fluid nodes while ensuring macroscopic conservation of mass and momentum. For turbulence simulations, a generalized slip-wall condition is applied and then corrected to match the target skin friction specified by the wall function. The original formulation was found to be overly diffusive, motivating refinements by Li [39] that improved accuracy. This method is implemented in the commercial solver PowerFLOW, where it has been successfully applied to a wide range of complex geometries and flow configurations using the very large-eddy Simulation (VLES) model based on the RNG $k-\varepsilon$ closure, yielding good agreement with experimental force coefficients and pressure distributions [1] [2] [3]. However, detailed validation of turbulence quantities in standard benchmark cases is scarce. The available studies comparing boundary-layer velocity profiles typically focus on flow separation regions and challenging physics scenarios using scale-resolving simulations [40] [41] [42] [43] [44].

Finally, in recent work Xue et al. [45] employed a physics-informed neural network to construct an LBM-based wall model, achieving very good results for turbulent channel flow across several friction Reynolds numbers. Their approach relies on a WMLES turbulence model combined with a boundary treatment that does not fall into any of the categories described above, as the near-wall effects are imposed through an external volumetric force applied to the first layer of cells adjacent to the wall.

**1.3 Objectives of the present study**

Most of the wall-boundary treatments discussed above have been evaluated primarily within the framework of scale-resolving turbulence simulations, where LBM's strengths, namely, its ability to capture unsteady flow phenomena with low numerical dissipation and high computational efficiency, are most apparent. These studies often involve canonical test cases such as turbulent channel flow or complex three-dimensional configurations featuring challenging flow physics, including smooth-surface separation. However, such simulations introduce additional layers of complexity that make it difficult to disentangle the intrinsic performance of the wall-treatment scheme from other numerical and modeling effects [46]. For this reason, scale-resolving simulations may not always provide the most convenient framework for assessing the standalone accuracy of wall-boundary treatments. In contrast, RANS-based simulations offer a more controlled and reproducible environment for code-to-code verification and validation. Therefore, testing the wall-boundary scheme within an LBM-RANS framework, through direct comparison with results from conventional RANS solvers, provides a valuable intermediate step before applying the wall treatment to high-fidelity turbulence models. Such comparisons help identify potential accuracy issues that stem from the LBM boundary formulation itself, rather than from the turbulence model. Furthermore, this verification step is essential for high-fidelity turbulence models that rely on RANS closures in the near-wall region, such as hybrid RANS/LES and VLES formulations.

In this context, a recent study by the authors [21] demonstrated that a slip-velocity bounce-back formulation, though rarely employed within the LBM-RANS framework, can deliver accurate predictions for two canonical turbulent flows: turbulent channel flow and the zero-pressure-gradient boundary layer. Building upon this work, the present study extends the slip-velocity bounce-back approach to handle curved geometries and high-Reynolds-number turbulent flows on multilevel adaptive Cartesian grids. To achieve this, we introduce several refinements to the wall-treatment methodology, aimed at improving the smoothness and accuracy of surface quantities such as pressure and skin-friction distributions. We show that the accuracy reported by [21] for canonical configurations is retained in more geometrically complex and aerodynamically relevant cases. The proposed wall-boundary treatment is rigorously assessed on two widely recognized aerodynamic benchmarks: the turbulent flow around a NACA0012 airfoil and the high-lift MD-30P30N configuration, both extensively used for turbulence model validation. In addition to global force coefficients and surface pressure/skin-friction distributions (typically reported in LBM studies), we present detailed analyses of boundary-layer velocity profiles and turbulent eddy-viscosity distributions, enabling a more comprehensive validation.

The remainder of this paper is organized as follows. Section 2 describes the numerical methodology, including the lattice Boltzmann formulation, turbulence model, wall-function treatment, and the proposed slip-velocity bounce-back approach for curved immersed boundaries on Cartesian grids. Section 3 presents a comprehensive validation of the method on two widely used aerodynamic benchmark cases, with detailed comparisons against reference RANS solutions and available experimental measurements. Section 4 summarizes the main findings and discusses potential avenues for future development and application of the proposed approach.

## 2. Numerical methods

### 2.1 Lattice Boltzmann solver

In this study, we employ an in-house lattice Boltzmann research code, LBMx that has been developed from scratch at INTA, to investigate numerical techniques for LBM in aerodynamic flow applications. The solver implements a standard second-order trapezoidal discretization of the lattice Boltzmann equation (LBE) in lattice units using the D2Q9 velocity model [27]

$$\left| f_i(\vec{x}+\vec{c}_i,t+1) \right\rangle = \left| f_i(\vec{x},t) \right\rangle - \mathbf{\Lambda}(\left| f_i(\vec{x},t) \right\rangle - \left| f_i^{eq}(\vec{x},t) \right\rangle) + (\mathbf{I}-\mathbf{\Lambda}/2)\left| F_i(\vec{x},t) \right\rangle \quad (1)$$

where $\left| f_i \right\rangle = [f_1, f_2, ...., f_Q]^T$ are the particle distribution functions colliding and streaming along the discrete lattice velocities $\vec{c}_i = [\left| c_{ix} \right\rangle, \left| c_{iy} \right\rangle]$ connecting the nodes on a regular square grid in 2D (cubic in 3D) where

$$\begin{aligned} \left| c_{ix} \right\rangle &= [0,1,-1,0,0,1,-1,1,-1]^T \\ \left| c_{ix} \right\rangle &= [0,0,0,1,-1,1,-1,-1,1]^T \end{aligned} \quad (2)$$

Here $\left| f_i^{eq} \right\rangle$ denotes the discrete equilibrium distribution, $\mathbf{\Lambda}$ is the collision operator, and $\left| F_i(\vec{x},t) \right\rangle$ accounts for external body forces, which are not considered in the present study.

The macroscopic fluid density $\rho$ and velocity $\vec{u}$ are obtained by taking statistical moments of the distribution functions:

$$\rho = \sum_i f_i, \quad \rho\vec{u} = \sum_i \vec{c}_i f_i \quad (3)$$

In this work, we employ a central moments-based (CM) collision operator for the LBE, following the formulation of de Rosis et al. [47]. The equilibrium distribution is expressed as a fourth-order expansion in a basis of Hermite polynomials $\mathbf{H}^{(n)}$

$$f_i^{eq} = w_i \rho \left[ 1 + \frac{\vec{c}_i \cdot \vec{u}}{c_s^2} + \frac{1}{2c_s^4} \mathbf{H}_i^{(2)} : \vec{u}\vec{u} + \frac{1}{2c_s^6}(\mathrm{H}_{ixxy}^{(3)} u_x^2 u_y + \mathrm{H}_{ixyy}^{(3)} u_x u_y^2) + \frac{1}{2c_s^8} \mathrm{H}_{ixxyy}^{(4)} u_x^2 u_y^2 \right] \quad (4)$$

where ":" denotes the Frobenius inner product (full contraction of the tensor, i.e. $H^{(2)}_{i\alpha\beta}u_\alpha u_\beta$), $w_1 = 4/9, w_{2...5} = 1/9, w_{6...9} = 1/36$ and $c_s = 1/\sqrt{3}$ is the lattice sound speed.

While the streaming step is performed in discrete velocity space, the relaxation (collision) is performed in central moment space, which requires transferring the populations between these two spaces. The corresponding transformation matrices are assembled as follows. First, lattice velocities are shifted by the local fluid velocity as

$$\begin{aligned}|\bar{c}_{ix}\rangle &= |c_{ix} - u_x\rangle \\ |\bar{c}_{iy}\rangle &= |c_{iy} - u_y\rangle\end{aligned} \quad (5)$$

where $\bar{c}_{i\alpha}$ denotes the shifted lattice velocities. Second, a suitable basis in CM space must be chosen. Here we adopt the non-orthogonal CM basis proposed by Rosis [48] for the D2Q9 model, which results in the following transfer matrix

$$\mathbf{T} = \begin{bmatrix} \langle \mathbf{c}_i |^0 \\ \langle \bar{c}_{ix} | \\ \langle \bar{c}_{iy} | \\ \langle \bar{c}_{ix}^2 + \bar{c}_{iy}^2 | \\ \langle \bar{c}_{ix}^2 - \bar{c}_{iy}^2 | \\ \langle \bar{c}_{ix}\bar{c}_{iy} | \\ \langle \bar{c}_{ix}^2 \bar{c}_{iy} | \\ \langle \bar{c}_{ix}\bar{c}_{iy}^2 | \\ \langle \bar{c}_{ix}^2 \bar{c}_{iy}^2 | \end{bmatrix} \quad (6)$$

where $\langle \square |$ denotes the row vector.

In terms of the central moments $|\bar{m}_i\rangle = \mathbf{T}|f_i\rangle$, the collision process is

$$|\bar{m}_i^*\rangle = (\mathbf{I} - \mathbf{K})|\bar{m}_i\rangle + \mathbf{K}|\bar{m}_i^{eq}\rangle \quad (7)$$

where $|\bar{m}_i^{eq}\rangle = \mathbf{T}|f_i^{eq}\rangle$ are the equilibrium CMs. $\mathbf{K}$ is the collision matrix in CM space, which is taken to be $\mathbf{K} = \text{diag}[1,1,1,\omega_b,\omega_\nu,\omega_\nu,1,1,1]$. The collision operator in discrete velocity space eq. (1) is then given by $\mathbf{\Lambda} = \mathbf{T}^{-1}\mathbf{K}\mathbf{T}$. To enhance computational efficiency, the mapping between the population and central-moment spaces is carried out using an auxiliary shift matrix, which maps raw moments to central moments [49]. Finally, it is noted that, due the structure of $\mathbf{K}$, the three highest-order moments are relaxed toward their equilibrium values during this step.

A Chapman-Enskog analysis of LBE with this collision model [48] recovers the weakly compressible athermal Navier-Stokes equations,

$$\begin{aligned}&\partial_t \rho + \partial_\alpha (\rho u_\alpha) = 0 \\ &\partial_t (\rho u_\alpha) + \partial_\beta (\rho u_\alpha u_\beta) = -\partial_\alpha p + \partial_\beta (\rho \bar{\nu}(\partial_\alpha u_\beta + \partial_\beta u_\alpha) + \rho(\bar{\nu}_b - \bar{\nu})\delta_{\alpha\beta}\partial_\gamma u_\gamma)\end{aligned} \quad (8)$$

where $p = \rho c_s^2$ is the pressure. The dimensionless kinematic and bulk viscosities, $\bar{\nu}$ and $\bar{\nu}_b$ are related to the relaxation frequencies appearing in the diagonal matrix $\mathbf{K}$ by

$$\omega_\nu = \left(\frac{\bar{\nu}}{c_s^2} + \frac{1}{2}\right)^{-1} \quad (9)$$

and

$$\omega_b = \left(\frac{\bar{\nu}_b}{c_s^2} + \frac{1}{2}\right)^{-1} \quad (10)$$

For turbulent flows modelled with RANS-based approaches, the dimensionless kinematic viscosity entering in the shear relaxation factor $\omega_\nu$ is replaced by the effective viscosity, $\bar{\nu}_{eff} = \bar{\nu} + \bar{\nu}_t$, where $\bar{\nu}_t$ is the dimensionless (in lattice units) turbulent viscosity predicted by the turbulence model.

As for the bulk viscosity, we adopt the same bulk relaxation parameter $\omega_b = 1$ originally proposed by de Rosis et al. [48]. As a result of this choice, a finite amount of bulk viscosity is implicitly introduced into the LBM formulation. While this additional bulk viscosity enhances numerical stability, particularly for high-Reynolds-number turbulent simulations, it also leads to undesirable damping of acoustic waves. Consequently, the present collision model, with the tuned parameters employed here, is not suitable for aeroacoustic applications.

LBMx employs multilevel Cartesian grids to accurately resolve complex geometries. The grids are generated using a dedicated in-house preprocessor designed to integrate seamlessly with the solver. This module utilizes an octree-based mesh generation strategy, incorporating several automated techniques [50]: Alternating Digital Trees (ADTs) for wall surface elements intersection probing, painting methods to distinguish fluid and solid regions, ray tracing to classify orphan nodes, and advancing-front techniques to construct boundary layers near wall geometries. By specifying the lattice resolution at wall surfaces, the resolution of the outer domain, and the number of cell layers between successive Cartesian levels, the preprocessor automatically builds a multilevel grid. It also generates the connectivity information required for inter-level data transfer and the link-wise data structure needed to enforce the turbulent wall boundary condition, as described in Section 2.3.

A cell-centered approach is used to manage grid information. Multilevel data transfer between grid levels follows the original Rohde algorithm [51], which is strictly mass-conservative across grid interfaces by avoiding intermediate-node interpolations used by other methods [52]. A recursive algorithm efficiently synchronize population transfers between grids levels, including the explosion/coalescence stages, the stream-collide procedure, and the turbulence model computations at each level [53] [54]. In simulations presented here, acoustic scaling is applied for time-step advancement, ensuring a constant Mach number across all grid levels. Each finer grid level advances with two sub-time steps per coarse-grid time step to maintain synchronization of population distributions. As a result, the computational cost is dominated by the finest grid level, highlighting the importance of employing near-wall modeling in LBM simulations of high-Reynolds-number turbulent flows.

**2.2 Turbulence modeling**

In this work, we employ the non-negative formulation of the Spalart-Allmaras turbulence model (SA-neg) [55] [56] to compute the turbulent eddy-viscosity, which is then used to modify the shear relaxation frequency in the LBM-RANS equations. This formulation is essential for stabilizing the solution when using the second-order finite-difference discretization of the the SA equation. The SA-neg model is specifically designed to prevent negative turbulent eddy-viscosity values, which can lead to numerical instabilities, particularly near boundary-layer edges and wake regions, where the turbulence transitions sharply over a short distances. In our simulations, spurious negative values were occasionally observed at the interfaces between different grid levels during the transient startup phase.

In the SA model, the eddy viscosity $v_t$ is expressed in terms of the auxiliary SA working variable $\tilde{v}$ as

$$v_t = \max(\tilde{v},0)f_{v1} \quad f_{v1} = \frac{\chi^3}{\chi^3 + c_{v1}^3} \quad \chi = \frac{\tilde{v}}{v} \qquad (11)$$

where $v$ is the molecular kinematic viscosity.

The modified turbulent viscosity $\tilde{v}$ obeys two transport equations depending on its sign. For $\tilde{v} \geq 0$ the original SA model with a modified vorticity to prevent negative values of $\tilde{S}$ is used

$$\partial_t \tilde{v} + u_\alpha \partial_\alpha \tilde{v} = P - D + \frac{1}{\sigma}\left[\partial_\gamma((v+\tilde{v})\partial_\gamma \tilde{v} + c_{b2}\partial_\gamma \tilde{v}\partial_\gamma \tilde{v})\right] \qquad (12)$$

where $\alpha, \gamma$ are Cartesian indices (and summation over repeated indices is understood), $u_\alpha$ are the Cartesian velocity components and

$$P = c_{b1}\tilde{S}\tilde{v}, \quad D = c_{w1}f_w\left(\frac{\tilde{v}}{d}\right)^2 \qquad (13)$$

are production and wall destruction terms. In eq. (13), $\tilde{S}$ is a modified vorticity defined to be positive

$$\tilde{S} = \begin{cases} \Omega + \bar{S}, & \bar{S} \geq -c_{v2}\Omega \\ \Omega + \dfrac{\Omega(c_{v2}^2\Omega + c_{v3}\bar{S})}{(c_{v3} - 2c_{v2})\Omega - \bar{S}}, & \bar{S} < -c_{v2}\Omega \end{cases}, \quad f_{v2} = 1 - \dfrac{\chi}{1 + \chi f_{v1}} \qquad (14)$$

where $\Omega = \sqrt{2W_{\alpha\beta}W_{\alpha\beta}}$, with $W_{\alpha\beta} = (\partial_\alpha u_\beta - \partial_\beta u_\alpha)/2$, is the magnitude of the vorticity, $d$ is the distance to the nearest wall and, and $f_w$ is the damping function:

$$f_w = g\left[\dfrac{1 + c_{w3}^6}{g^6 + c_{w3}^6}\right]^{1/6}, \quad g = r + c_{w2}(r^6 - r), \quad r = \min\left(\dfrac{\tilde{v}}{\tilde{S}\kappa^2 d^2}, r_{\lim}\right) \qquad (15)$$

For negative values $\tilde{v} < 0$, the following alternative transport equation is employed

$$\partial_t \tilde{v} + u_\alpha \partial_\alpha \tilde{v} = P_n - D_n + \dfrac{1}{\sigma}\left[\partial_\gamma((v + \tilde{v}f_n)\partial_\gamma \tilde{v}) + c_{b2}\partial_\gamma \tilde{v}\partial_\gamma \tilde{v}\right] \qquad (16)$$

with

$$P_n = c_{b1}(1 - c_{t3})\Omega\tilde{v}, \quad D_n = -c_{w1}\left(\dfrac{\tilde{v}}{d}\right)^2, \quad f_n = \dfrac{c_{n1} + \chi^3}{c_{n1} - \chi^3} \qquad (17)$$

Finally, the constants of the model are

$$\begin{aligned} &c_{b1} = 0.1355, \quad c_{b2} = 0.622, \quad \sigma = 2/3, \quad \kappa = 0.41 \\ &c_{w1} = c_{b1}/\kappa^2 + (1 + c_{b2})/\sigma, \quad c_{w2} = 0.3, \quad c_{w3} = 2.0 \\ &c_{v1} = 7.1, \quad r_{\lim} = 10.0, c_{v2} = 0.7, c_{v3} = 0.3, \ c_{t3} = 1.2, \\ &c_{n1} = 16.0 \end{aligned}$$

The model constants and formulation follows the recommendation in [56] and were selected to facilitate comparison with reference data in National Aeronautics and Space Administration (NASA)'s Turbulence Modeling Resource website [57]. The turbulence model equation is discretized using a second-order finite-difference method. Velocity gradients required for vorticity computation are reconstructed via an unweighted least-squares (LSQ) method [58], using the velocities of the nearest nodes. The convective term is discretized with a second-order Lax-Wendroff (LW) scheme[1], while diffusive terms are approximated using standard second-order centered finite differences. Since second-order accuracy requires a three-point stencil in each coordinate direction, all finite-difference operators described above are fully defined at fluid nodes (F) adjacent to boundary nodes (see Figure 1).

---

[1] In addition to the second order convective term discretization, a first order upwind scheme has also been implemented to assess the influence of the discretization order on the results. A comparison of both approaches is presented in section 3.3.

Turbulent viscosity values at interface-grid nodes on each grid level are interpolated using the LSQ method, based on interface information between grid levels. The Lax-Wendroff scheme may introduce spurious oscillations at interfaces between grid regions with different resolution levels, particularly during the initial transient phase; these oscillations usually dissipate as the simulation progresses. Although more advanced second-order convective discretizations, such as total variation diminishing (TVD) schemes with flux limiters, could reduce these oscillations, they were not employed in order to minimize the computational cost per time step.

Boundary conditions are prescribed as follows: at wall boundary nodes (see Figure 1), a Dirichlet condition is used for the turbulence variable $\tilde{v}_B = v\kappa d_{WB}^+$, where $d_{WB}$ denotes the distance of this node from the wall. This condition follows from the construction of the SA model that admits the solution $\tilde{v} = \kappa u_\tau y = \kappa v y^+$ in accordance with the law of the wall, where $u_\tau$ is the friction velocity, $\kappa$ is the von Karman constant and $y^+ = y u_\tau / v$. Moreover, the equilibrium wall function employed in this work (described below) is constructed to satisfy this solution across all three regions of the law of the wall (viscous sublayer, buffer layer and logarithmic layer). Therefore, the prescribed value of $\tilde{v}_B$ may be applied, in principle, regardless of the precise location of the boundary node. At the domain inlet, the turbulent viscosity is prescribed as $\chi = 3.0$, while at the domain outlet a first-order extrapolation from interior values is employed.

For the wall function, we employ an analytical solution for the velocity that is consistent with the SA model in the law-of-the-wall region. A detailed derivation of this solution is provided in [56]. The derivation of this wall function assumes that both convection and pressure-gradient terms are negligible, and $\partial_x \ll \partial_y$, where $x$ and $y$ denote the streamwise and wall-normal directions respectively. Under these assumptions, the streamwise momentum equation reduces to the turbulent boundary-layer diffusion equation

$$\frac{d}{dy}\left[(v+v_t)\frac{du}{dy}\right] = 0 \qquad (18)$$

which can be integrated to yield

$$(v+v_t)\frac{du}{dy} = u_\tau^2 \qquad (19)$$

Reference [56] shows that substituting this velocity gradient into the SA model equation leads to the following solution for the model variable,

$$\chi = ky^+$$
$$\tilde{S} = \left(\frac{u_\tau^2}{\nu}\right)\left(\frac{1}{ky^+}\right) \tag{20}$$

from where the dimensionless velocity gradient can be expressed as

$$\frac{du^+}{dy^+} = \frac{c_{v1}^3 + (ky^+)^3}{c_{v1}^3 + (ky^+)^3(1+ky^+)} \tag{21}$$

Integration of this expression provides the analytical form of the wall function:

$$u_{WF}^+(y^+) = \bar{B} + c_1 \log\left((y^+ + a_1)^2 + b_1^2\right) - c_2 \log\left((y^+ + a_2)^2 + b_2^2\right)$$
$$- c_3 \arctan\left(b_1/(y^+ + a_1)\right) - c_4 \arctan\left(b_2/(y^+ + a_2)\right) \tag{22}$$

where the model constants take the following values:

$$\bar{B} = 5.033908790505579$$
$$a_1 = 8.148221580024245, \quad a_2 = -6.9287093849022945$$
$$b_1 = 7.4600876082527945, \quad b_2 = 7.468145790401841$$
$$c_1 = 2.5496773539754747, \quad c_2 = 1.3301651588535228$$
$$c_3 = 3.599459109332379, \quad c_4 = 3.639753186868684494$$

It should be emphasized that this wall function is derived under the assumption of equilibrium flows. Consequently, its use in aerodynamic flows should be limited to conditions that remain close to turbulent equilibrium.

### 2.3 Slip-velocity bounce-back wall treatment for LBM-RANS

We extend the slip-velocity bounce-back boundary scheme of [21], which was originally developed for LBM-RANS computations over flat, grid-aligned surfaces, to handle curved surfaces immersed in Cartesian grids. The slip velocity prescribed in this work is derived from an analysis of how the bounce-back condition, augmented with the wall momentum correction term, enforces the macroscopic flow boundary condition at the boundary node. As argued in [21], the scheme enforces the turbulent boundary-layer diffusion model [59] [60] at the boundary node B (see Figure 1)

$$\left[(\nu + \nu_t)\partial_y u\right]_B = u_\tau^2 \tag{23}$$

For flat, grid-aligned surfaces, the half-way slip-velocity bounce-back condition

$$f_i(\vec{x}_b, t+1) = f_i^*(\vec{x}_b, t) - 2w_i \rho_w \frac{\vec{c}_i \cdot \vec{u}_w}{c_s^2} \qquad \vec{c}_i \cdot \vec{n} > 0 \tag{24}$$

can be shown, via simplified second-order Chapman-Enskog analysis [27]), to approximate the following macroscopic boundary condition

$$u_B = u_W + \partial_y u \big|_B d_{WB} + O((\tau - 1/2)^2 \partial_y^2 u \big|_B) \qquad (25)$$

The associated truncation error depends on the dimensionless relaxation time ($\tau = \omega^{-1}$) and remains small for high Reynolds number turbulent flows [21].

Now if the slip velocity $u_W$ is defined as

$$u_W = u_B - \frac{u_\tau^2}{\nu + \nu_t \big|_B} d_{WB} \qquad (26)$$

it follows that the diffusion model at node B (23) is implicitly satisfied, $(\nu + \nu_t)_B \partial_y u \big|_B \approx u_\tau^2$, as can be easily seen by inserting (26) into (25).

The key feature of the boundary scheme formulation proposed in [21] for LBM-RANS computations is the estimation of the velocity $u_B$ at the boundary node using an analytical wall function. First, the friction velocity $u_\tau$ is obtained by solving (22) with the wall-parallel velocity $u_R$ sampled at a location R from the wall. Then, $u_B$ is evaluated from the analytical wall function (22) using the previously determined $u_\tau$. Once the slip-velocity $u_W$ is obtained with (26), the incoming distribution functions at the boundary node are reconstructed using (24), which is however only strictly valid for the case that the wall coincides with a cell boundary. In the general case, eq. (24) is replaced with Yu's interpolated bounce-back scheme [61]. In this scheme, the missing distribution functions are evaluated as

$$f_i(x_W, t+1) = f_i(x_R, t+1) + (1-q_i)f_i^*(x_B, t) - 2\rho_w w_i \frac{\vec{c}_i \cdot \vec{u}_w}{c_s^2}$$

$$f_i(x_B, t+1) = f_i(x_W, t+1) + \left(\frac{q_i}{q_i+1}\right)\left(f_i(x_R, t+1) - f_i(x_W, t+1)\right) \qquad (27)$$

where $f_i^*$ are post-collision distributions, $q_i = |\vec{x}_B - \vec{x}_{W_i}|/|\vec{c}_i \Delta t|$, and $\vec{x}_{W_i}$ represents the intersection point of the corresponding lattice link with the wall. The reader is referred to the original publication [61] for a detailed derivation.

The slip-velocity bounce-back approach described above has shown to yield accurate predictions for turbulent channel flow and zero-pressure-gradient turbulent boundary layer over a flat plate within the LBM-RANS framework [21]. A key distinction from the formulation proposed by Nishimura et.al. [33] lies in their treatment of the slip-velocity: rather than using information from node B through the wall function, they directly employ the interpolated velocity from the flow field at the exchange location R, which is situated relatively far from the wall. While this strategy is common in wall-modeled LES formulations based on the equilibrium wall model, it can introduce significant errors in an LBM-RANS context, because it neglects the near-wall information provided by the wall-function model.

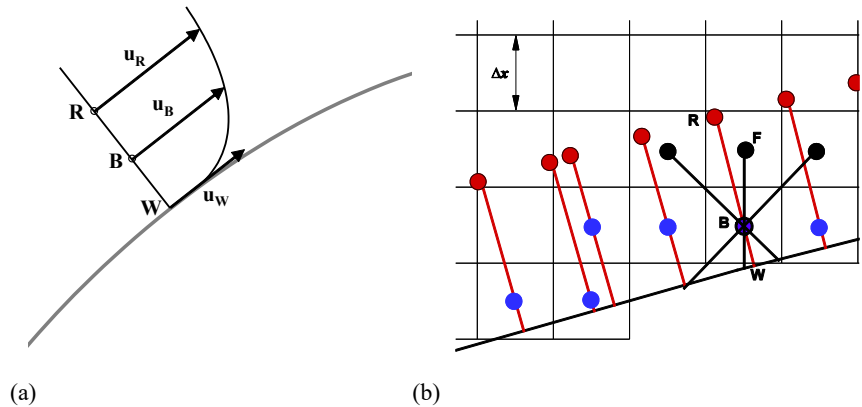

(a)                                    (b)

Figure 1. (a) Schematic illustration of the information required to apply the slip-velocity bounce-back scheme for turbulent flows with wall functions. (b) Illustration of the challenge in Cartesian meshes: although reference points (red dots) are defined at a fixed distance from the wall, the boundary nodes (blue dots) vary their distance rapidly over only a few lattice spacings (from zero to approximately $O(\Delta x)$). Links used in the bounce-back scheme at node B are illustrated with black lines.

Extending this method to curved surfaces introduces two primary challenges. First, unlike the case of flat, grid-aligned surfaces, the reference node R no longer coincides with a fluid node. Consequently, the flow variables, density and velocity, must be reconstructed at R through a robust and accurate interpolation procedure. The second challenge, which becomes particularly critical when wall modeling is employed, stems from the rapid variation of wall distance associated with boundary nodes in immersed Cartesian grids (see Figure 1(b)). In such cases, the wall distance $d_{WB}$ may vary by $O(\Delta x)$, where $\Delta x$ is the local grid spacing as illustrated in Figure 1(b), or even reach zero within only a few lattice spacings. Directly applying (26) under these conditions can lead to large oscillations in the computed slip velocity, resulting in noisy wall-shear predictions. To mitigate this issue, we introduce an auxiliary virtual node V (see Figure 2), positioned along the wall-normal direction between the wall intersection point IW and the reference node R. This virtual node is placed at a prescribed, fixed distance $d_{WV}$ from the wall, ensuring a smoother and more consistent evaluation of the slip velocity and reducing spurious oscillations in the near-wall solution.

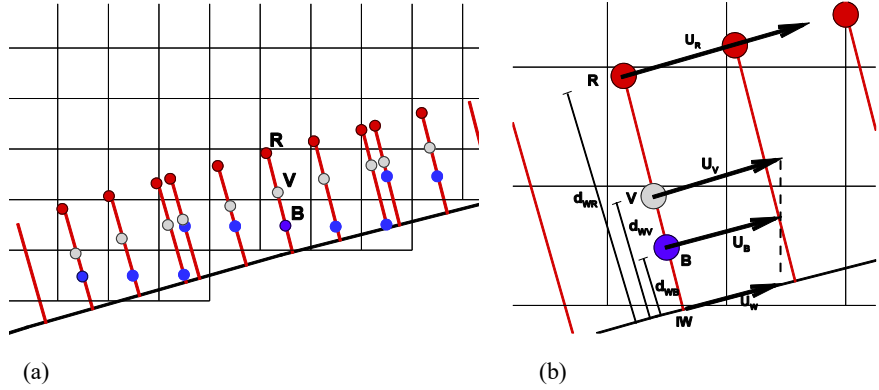

(a)  (b)

Figure 2. Virtual-node concept (a) boundary nodes (B) are colored in blue, reference nodes for probing are colored in red and virtual-nodes are colored in gray. (b) Detail of the information required to apply the slip-velocity bounce-back scheme for turbulent flows with wall functions.

Assuming a linear variation of velocity with wall distance (see Figure 2 (b)), the wall-normal velocity gradients at the virtual V and the boundary node B are considered identical, i.e., $\partial_y u|_V = \partial_y u|_B$. Under this assumption the slip-velocity is first computed from the diffusion equation in a manner analogous to Eq. (26), but using the virtual-node velocity $u_V$ instead of $u_B$

$$\left[(\nu+\nu_t)\partial_y u\right]_V \approx (\nu+\nu_t)|_V (u_V - u_W)/d_{WV} = u_\tau^2 \quad \Rightarrow \quad u_W = u_V - \frac{u_\tau^2}{(\nu+\nu_t)_V} d_{WV} \qquad (28)$$

where $u_V$ is obtained from the analytical wall function. The slip-velocity bounce-back scheme (27) then adjusts implicitly the velocity at node B such that the resulting wall-normal gradient satisfies the turbulent diffusion model at the virtual node V. This procedure ensures a smooth and physically consistent near-wall velocity profile while significantly reducing spurious oscillations associated with abrupt variations in wall distance.

The implementation of this boundary scheme, hereafter referred as the Virtual Node (VN)-slip-velocity bounce-back for conciseness, is straightforward. For each boundary node B (see Figure 2 (b)), a wall-normal line passing through node B is constructed during the preprocessing stage. The intersection of this line with the corresponding surface element, segments in 2D or triangles in 3D, is identified as point IW. Along this line, the reference point R and the virtual node V are positioned at prescribed distances $d_{WR}$ and $d_{WV}$, respectively, from the wall. Based on preliminary tests, distances of $d_{WR} = 2\Delta x$ and $d_{WV} = \Delta x$, provide robust results while keeping the reference nodes (R) outside the boundary cells. The steps for implementing the boundary-wall scheme are summarized as follows:

1) Interpolate density and velocity at R: Density is obtained via inverse distance weighting (IDW) [20] [22] using nearest neighbors excluding boundary nodes to reduce pressure oscillations, and velocity is computed using a least-squares (LSQ) approach.
2) Compute the tangential velocity component as $\vec{u}_{R_t} = \vec{u}_R - (\vec{u}_R \cdot \vec{n})\vec{n}$ where $\vec{n}$ is the unit normal vector to the surface.
3) Determine the friction velocity $u_\tau$ by solving (22) at R with a Newton-Raphson method.
4) Evaluate the velocity at the virtual-node V using the analytical wall function $u_V = u_\tau f_{SA}(y_V^+)$
5) Compute the slip-velocity $\vec{u}_W = u_W \vec{u}_R / |\vec{u}_R|$ where $u_W$ is given by (28) with $(\nu_t)_V = \nu \kappa y_V^+$
6) Reconstruct the missing distribution functions at node B using Yu's interpolated bounce-back scheme, assuming negligible density variation along the wall-normal direction, i.e. $\rho_W \approx \rho_R$.

The procedure described above is similar to other approaches developed to enforce turbulent wall boundary conditions with wall modeling in LBM. In particular, steps 1 and 2 are generally required for immersed wall boundary conditions, with most differences arising from the interpolation techniques employed. Within the LBM framework, the fundamental distinction occurs in step 6, as discussed in Section 1.

In regularized-based approaches, all distribution functions at boundary nodes are replaced by the sum of their equilibrium and non-equilibrium components. The non-equilibrium part is typically reconstructed using estimates of the velocity gradients [20] [22] [23]. Consequently, in these methods, steps 4 – 6 are replaced by the reconstruction of the equilibrium and non-equilibrium distribution functions.

In contrast, compared to previously proposed bounce-back approaches (specifically those based on Nishimura's approach [33] [34]) the introduction of the new steps 4 and 5 is critical for improving the accuracy of bounce-back-based boundary conditions within the LBM-RANS framework, as will be demonstrated in Section 3.1.5

## 3. Results and discussion

The accuracy of the proposed VN-slip-velocity bounce-back scheme is assessed on two high-Reynolds-number turbulent flows past a NACA0012 and a MD-30P30N multi-element airfoil. These cases were previously simulated using LBM-RANS with regularized wall boundary conditions with the proLB code [7] [22]. The NACA0012 case is used first to validate the scheme against standard finite volume RANS solutions, comparing surface quantities such as pressure and skin-friction coefficients, as well as detailed flow field data including turbulent boundary layer profiles and turbulent viscosity distributions. The MD-30P30N configuration is then employed to validate the approach for a more geometrically and physically complex aerodynamic scenario. Overall, these tests demonstrate that the ability of the proposed LBM methodology to reproduce standard RANS results in practical aerodynamic applications.

### 3.1 NACA 0012 turbulent flow

#### 3.1.1 Test case description and computational setup

The NACA 0012 airfoil is first employed to verify and validate the proposed boundary-condition scheme. This aerodynamic test case has been recommended for verification and validation activities on the Turbulence Model Resource (TMR) website [57], where reference solutions are provided for code-to-code comparison. In particular, solutions obtained with the NASA CFL3D code using the SA-neg model on a highly refined mesh with near-wall resolution ($y^+ < 1$) are available. The flow is essentially incompressible (Mach number of 0.15) with a Reynolds number of $\text{Re} = 10^6$ based on the airfoil chord. Surface quantities, including pressure coefficient and skin-friction distributions, are available at the TMR website at angles of attack of 0º, 10º, and 15º. In addition to these reference solutions, RANS computations have been carried out using Fluent (version 2024-R2). The results of these simulations are used for detailed flow verification in Sections 4.1.2 and 4.1.3. The Fluent mesh is structured-like and features a highly refined boundary-layer region generated according to standard best practices.

Although the TMR web resource identifies the 10° angle of attack as the primary reference for comparison, this study focuses particularly on the 0° case. At this angle, the upper boundary layer develops under only mild favorable and adverse pressure gradients near the leading edge and a very weak adverse pressure gradient along the remainder of the airfoil. Under these conditions, the equilibrium wall-function assumption, on which the analytical wall function is based, is nearly satisfied over the entire surface. This allows us to isolate and analyze the effect of the geometrical curvature of the airfoil, independently of other physical influences such as pressure-gradient effects on the turbulent boundary layer. Nevertheless, the 10° angle of attack, which was selected by TMR and the 6[th] American Institute of Aeronautics and Astronautics (AIAA) CFD drag prediction workshop [62] (DPW6) as case 1 for verification studies, is also considered here. This configuration involves stronger flow acceleration and deceleration near the leading edge and, consequently, larger deviations of the LBM results (which are based on an equilibrium wall model) from the reference solutions are expected. Accounting for these effects would require an improved wall model including non-equilibrium terms, as discussed by Berger et. al. [59].

The NACA 0012 airfoil geometry provided by the TMR website is used in this study. Reference chord is defined as $c_{ref}$ = 1m. Four grids were generated automatically by prescribing the surface resolution using the Octree-Cartesian mesh preprocessor of LBMx. Table 1 summarizes the main features of the meshes for an angle of attack of 0º. The computational domain extends to $[L_x/c_{ref}, L_y/c_{ref}] = [100,100]$ in order to minimize outer boundary influence while maintaining computational efficiency. Figure 3 shows a view of the three meshes for an angle of attack of 10º in a region near the airfoil leading edge.

Table 1: Mesh parameters used in the grid-convergence study of the NACA0012 airfoil at $\alpha$ = 0º. The second column reports the lattice resolution at the airfoil surface, and the third column lists the total number of elements/lattices in the corresponding mesh. The last column provides the corresponding computed $y_V^+$ values for the virtual nodes at mid-chord. (*) CFL3D grid information is taken from https://turbmodels.larc.nasa.gov/.

| Mesh | $\Delta x_{min}/c$ | Nelem | Octree levels | $y_V^+$ (x/c = 0.5) |
|---|---|---|---|---|
| coarse | 1.5×10⁻³ | 45340 | 11 | 352 |
| medium | 7.6×10⁻⁴ | 81121 | 12 | 174 |
| fine | 3.8×10⁻⁴ | 152837 | 13 | 92 |
| ultra-fine | 1.9×10⁻⁴ | 297243 | 14 | 47 |
| CFL3D (*) | -- | 230529 (897×257) | -- | <1 |
| Fluent | -- | 115136 (448×128) | -- | <1 |

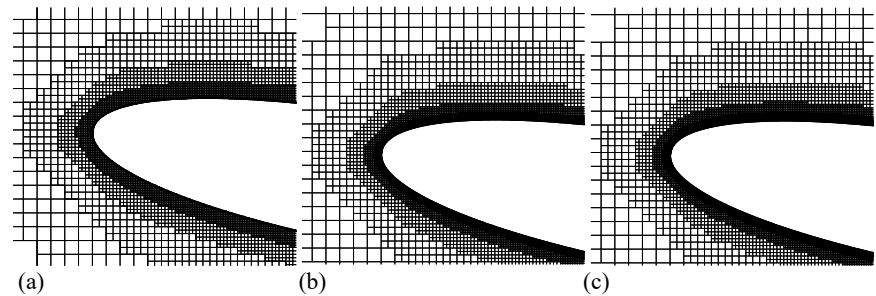

Figure 3. Illustration of the three mesh resolutions used for the NACA0012 validation and verification test case: (a) coarse mesh. (b) medium mesh, and (c) fine mesh

The LBM boundary conditions at the domain boundaries are as follows: inlet velocity is prescribed at the domain entrance using the Zou-He method [63]; the top and bottom surfaces are treated as symmetry planes, while the outlet uses an extrapolation method based on Yu's approach [64]. These boundary conditions are applied consistently across all test cases in this study. The inlet velocity is set to $U_{ref}$ = 90 m/s, the reference density $\rho_{ref}$ = 1 Kg.m⁻³, and the physical kinematic viscosity $\nu$ = 1.5×10⁻⁵ m²s⁻¹. The Reynolds number based on the reference chord is $Re_{c_{ref}}$ = 6×10⁶, and the Mach number is $M$ = 0.15. LBM simulations are performed using acoustic scaling to define the physical time step, and fully turbulent simulations are conducted. The inlet modified turbulence is prescribed as $\tilde{\nu}/\nu = 3.0$ and the turbulent field is initialized with the same value. Simulations are run until convergence of drag and lift coefficients (defined by Eq. (30) in Section 3.1.3), which typically requires 30-50 convective time units. Figure 4 presents the time history of the drag and lift coefficients for the fine grid at angles of attack of 0º and 10º. For $\alpha = 10º$, the relative variations of the coefficients over the last five convective time units are $\Delta C_{Df}/C_{Df} < 3\times10^{-4}$, $\Delta C_D/C_D < 1.5\times10^{-2}$ and $\Delta C_L/C_L < 5\times10^{-3}$, which are representative of the convergence criterion adopted in this study. For $\alpha = 0º$, the drag variation is $\Delta C_D/C_D < 2\times10^{-4}$ for both drag components.

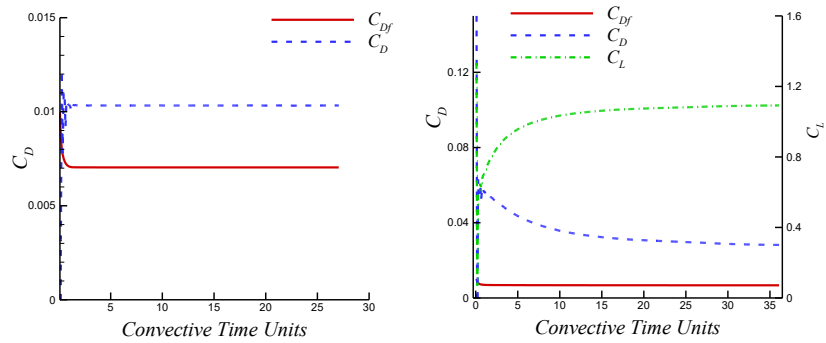

Figure 4. Time history of the lift and drag coefficients of the NACA0012 airfoil expressed in Convective Time Units ($t/(c_{ref}/U_{ref})$). (a) $\alpha = 0º$; (b) $\alpha = 10º$

### 3.1.2 Surface data results

Figure 5 plots the pressure and friction coefficient distributions obtained using the VN bounce-back boundary condition for the fine-resolution grid ($\Delta x_{min} = 3.8\times10^{-4}$, $y^+ = 100$ at mid chord).These results are compared with the reference CFL3D RANS solutions provided by NASA and with Fluent RANS computations employing the standard Spalart-Allmaras (SA) model.

The pressure and friction coefficients are defined, respectively, as

$$C_p = \frac{2(p - p_{ref})}{\rho_{ref} U_{ref}^2} \quad , \quad C_f = 2\left(\frac{\rho_w}{\rho_{ref}}\right)\left(\frac{u_\tau}{U_{ref}}\right)^2 \tag{29}$$

For the $\alpha = 0º$ case, Figure 5(a) shows that the pressure distribution predicted by the LBM approach is in good overall agreement with the reference RANS solutions. The corresponding skin-friction distribution, shown in Figure 5(b), also exhibits reasonably good agreement, particularly in the aft region of the airfoil ($x/c \geq 0.45$), where the influence of the coarser resolution near the leading edge on the LBM is less significant. The RANS solutions employ a highly refined mesh near the leading edge, whereas the present LBM approach uses a Cartesian mesh that does not fully resolve the leading edge turbulent boundary layer when combined with a wall-function treatment. Additional refinement in this region would likely improve the accuracy of both pressure and skin-friction predictions (see section 3.1.3). Notably, the VN-bounce-back scheme provides comparatively smooth profiles for both quantities, which is a desirable property for immersed-boundary methods based on Cartesian grids [22] [65].

The results for $\alpha = 10°$ are presented in Figure 5(c) and (d) and are compared against results obtained with Fluent, since TMR data does not include the skin friction distribution for the lower side of the airfoil. These figures indicate that the present approach maintains generally good agreement even in the presence of stronger pressure gradients near the leading edge. The suction peak in the pressure distribution is slightly underpredicted, and the skin-friction coefficient is somewhat overpredicted in the leading edge region, likely due to the limited grid resolution in that area. Streamwise-varying grid refinement would likely improve the prediction of suction peaks and of the boundary layer development, although this feature is not yet available in the current version of the LBMx solver. It should also be noted that grid refinement alone may not fully reproduce the resolved-turbulence CFL3D solutions, as wall-function modeling errors may contribute to the observed discrepancies. Incorporating an ODE-based wall model could mitigate these effects [59] [66].

Overall, and despite the unveiled limitations, the proposed wall-boundary treatment yields reasonably good predictions of both pressure and skin-friction distributions, particularly in regions where the turbulent boundary layer is sufficiently resolved by the wall function approach.

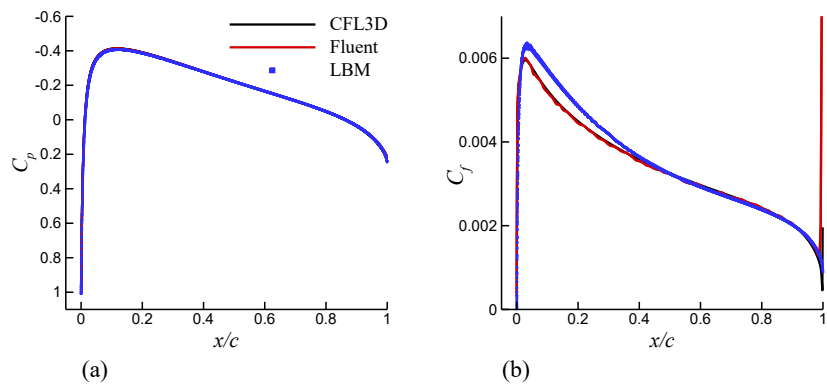

(a)        (b)

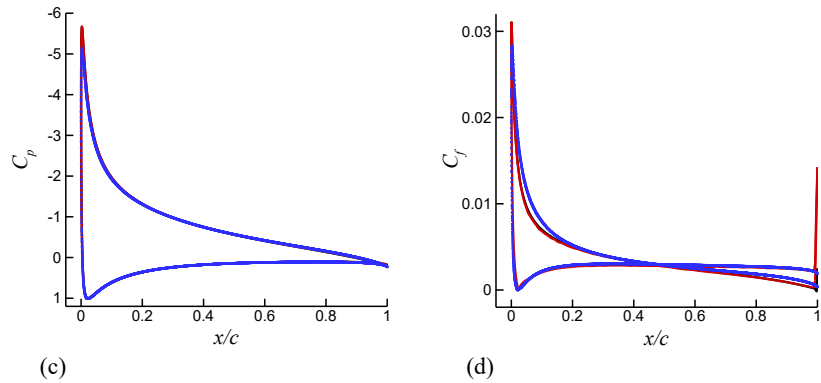

(c)                                                           (d)

Figure 5. Surface data obtained with the VN-slip-velocity bounce-back scheme on the fine grid, compared with the reference CFL3D solution *[57]* and Fluent V18.2 results at $M = 0.15$ and Re = $6 \times 10^6$. (a) Pressure-coefficient distribution at $\alpha = 0°$, (b) skin-friction coefficient distribution at $\alpha = 0°$, (c) pressure-coefficient distribution at $\alpha = 10°$, and (d) skin-friction coefficient distribution at $\alpha = 10°$. LBM results are shown as blue symbols, while CFL3D and Fluent solutions are represented by black and red lines, respectively.

### 3.1.3 Grid sensitivity study

A grid-convergence study was conducted to assess the robustness and sensitivity of the proposed wall treatment boundary condition to grid resolution. Cartesian meshes were generated by successively halving the lattice spacing prescribed at the wall boundary. The main characteristics of the meshes are summarized in Table 1. The so-called ultrafine mesh represents the highest resolution compatible with the equilibrium wall function approach under uniform grid refinement, assuming that the first lattice node remains within the logarithmic layer. Employing finer resolutions would cause a significant fraction of the virtual nodes (V-nodes) to fall within the buffer layer in the aft region of the airfoil ($x/c > 0.5$), which could reduce the accuracy of the wall function model. Moreover, such resolutions would be computationally prohibitive for three-dimensional simulations due to the resulting mesh size. A comparison of the $y^+$ distribution for the R, B and V nodes is presented in Figure 6 for the four grid resolutions considered. The results show a dispersion of the $y+$ values at the boundary nodes (B) due to their varying distances from the wall, ranging from 0 to a distance of the order of $\Delta x$ for cell-centered Cartesian grids. In contrast, the V nodes are consistently located within the logarithmic layer, where the turbulent equilibrium assumption is satisfied.

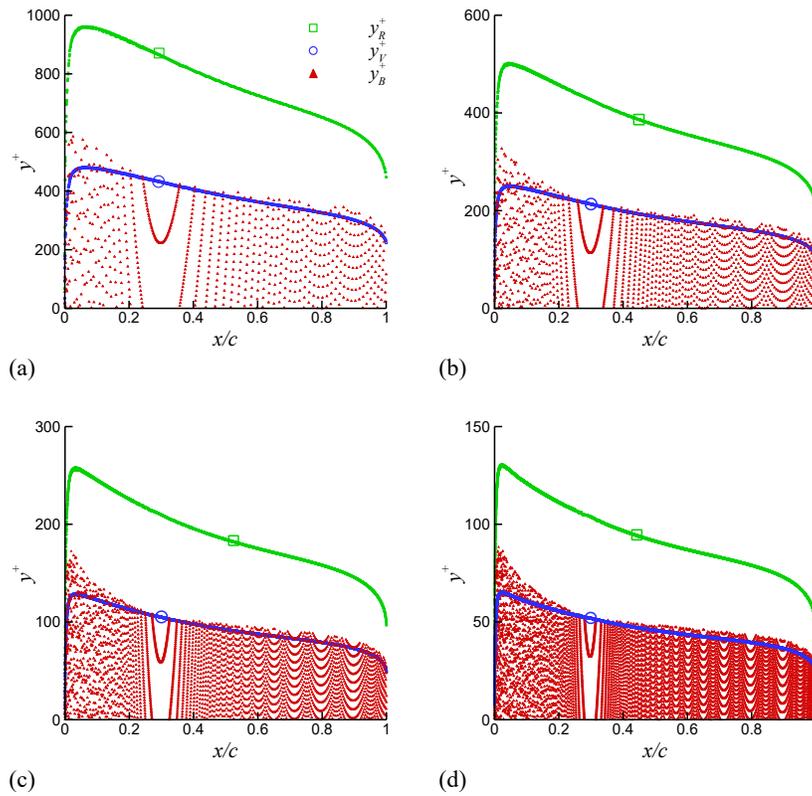

Figure 6. $y^+$ distribution for the R, B and V nodes across the four mesh resolutions. (a) coarse mesh. (b) medium mesh. (c) fine mesh. (d) ultra-fine mesh. Note that virtual nodes (V) are consistently located within the logarithmic layer for all resolutions.

Figure 7 compares the pressure and skin-friction coefficient distributions obtained for the different grid resolutions with the reference CFL3D solution.

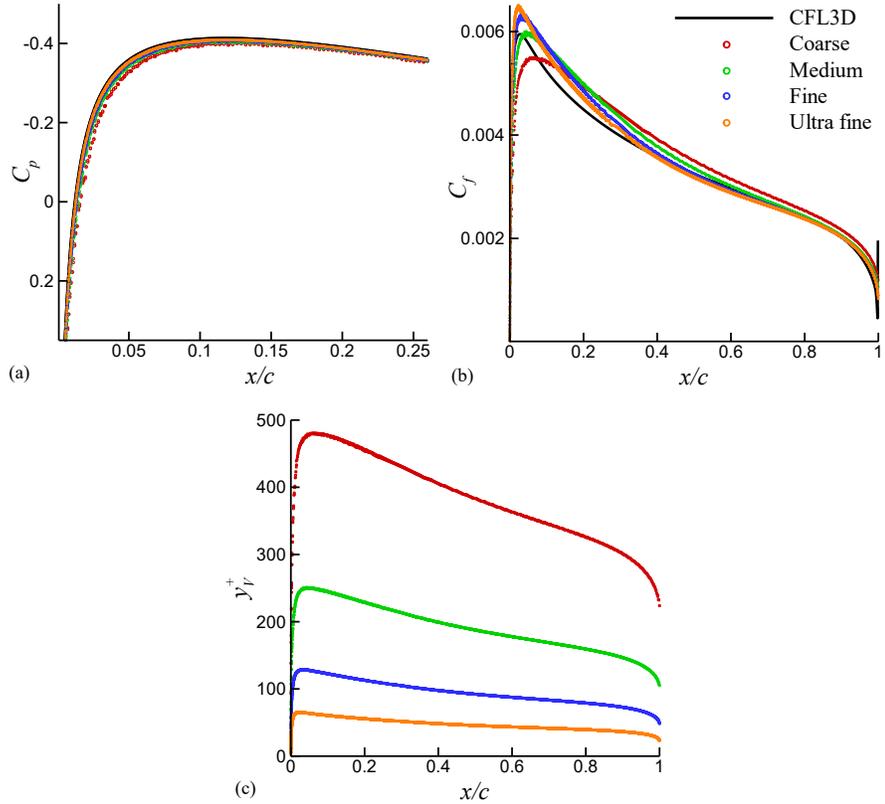

Figure 7. Grid convergence study for $\alpha = 0°$. (a) Detail of pressure coefficient ($C_p$) distribution near the leading-edge, (b) skin-friction coefficient $C_f$ distribution along the airfoil surface, (c) $y_V^+ = (u_\tau y_V)/\nu$ distribution along the airfoil. Red symbols: coarse grid, green symbols: medium grid, blue symbols: fine grid and orange symbols: ultra-fine grid.

The results indicate that increasing the grid resolution consistently reduces the discrepancies between the LBM and reference solutions. Interestingly, the skin-friction coefficient shows very little sensitivity to grid resolution (excluding the leading-edge effect), a trend also reported in *[21]*. This behavior may be advantageous compared to other LBM turbulence wall-treatment approaches, in which the friction coefficient seems to exhibit greater sensitivity to lattice resolution. On the down side, discrepancies in the skin-friction coefficient near the leading edge (*x/c* < 0.2) increase with grid refinement. This behavior has been reported previously for the same test case using Cartesian grids with finite volume RANS solvers (see e.g. *[67]*) and in LBM-RANS grid refinement studies for the flat plate test cases ( *[22] [23]*). The increase in discrepancy can be attributed to the fact that the reference nodes (R) near the leading edge are located outside the developing boundary layer, where the local boundary-layer thickness is very small. In this region, the estimation of the friction velocity using a wall function becomes inaccurate, leading to an overprediction of the friction velocity. As the grid is refined, reference nodes are positioned closer to the wall but still outside the turbulent boundary layer edge. Consequently, the only way to enforce the logarithmic law of the wall is to increase the predicted friction velocity. This explains the observed trend, which reflects the interaction between node placement and the actual boundary-layer development in the leading-edge region.

The lift and drag coefficients were obtained by numerically integrating the pressure and skin-friction coefficient distributions over the airfoil surface as follows

$$\vec{C}_F = \frac{1}{c_{ref}} \sum_{S_k} (C_{f,sign})_k \vec{t}_k ds_k \quad , \quad \vec{C}_P = -\frac{1}{c_{ref}} \sum_{S_k} (C_p)_k \vec{n}_k ds_k$$
$$C_{Df} = \vec{C}_F \cdot \vec{e}_x \quad , \quad C_{Dp} = \vec{C}_p \cdot \vec{e}_x \quad , \quad C_D = C_{Df} + C_{Dp}, \quad C_L = \vec{C}_p \cdot \vec{e}_y$$
(30)

where $\vec{n}_k$ and $\vec{t}_k$ are the normal and tangential vectors of the *k-th* surface element (with the normal vector pointing into the fluid), $C_{f,sign}$ is given by eq. (29) with a sign correction depending on the relative orientation of the local velocity and the surface tangent, and $\vec{e}_x$ is the unit vector in the freestream direction. A dedicated algorithm was implemented to locate the closest boundary nodes of each surface element centroid and interpolate the corresponding pressure and friction coefficients to the centroid location.

The results are summarized in Table 2 and Table 3, which also include Fluent results for comparison. The errors relative to the reference CFL3D solution decrease consistently with grid refinement, indicating improved accuracy of the proposed approach. A nearly linear dependence of the pressure-drag component on the minimum lattice spacing is observed for both angles of attack. For $\alpha$ = 0º, the skin-friction drag varies by less than 3% between the coarse and ultrafine meshes, suggesting that the wall-boundary treatment is numerically consistent. Moreover, the friction drag exhibits only weak sensitivity to grid resolution for both angles of attack, in contrast to the stronger resolution dependence reported for other LBM turbulence-wall treatments (e.g., PowerFLOW results in [68]).

Finally, the difference in skin-friction drag between the ultrafine-mesh LBM results and the Fluent reference solution for the *α* = 0º case is below 1% and approximately 1.2 % relative to the value reported in [69] using the Universal Velocity Profile (UVP) integral method. For the *α* = 10º configuration, the discrepancy with respect to the CFL3D solution is about 9%, further supporting the robustness of the proposed boundary-condition formulation.

Table 2: Summary of the global aerodynamic coefficients computed with LBM for different grid resolutions at $\alpha = 0º$. CFL3D values are taken from the NASA turbulence modeling resource (https://turbmodels.larc.nasa.gov/)

| Mesh | $C_{Df}$ | $C_{Dp}$ | $C_D$ | $C_L$ |
| --- | --- | --- | --- | --- |
| coarse | $7.187 \times 10^{-3}$ | $6.149 \times 10^{-3}$ | $1.333 \times 10^{-2}$ | $-1.238 \times 10^{-5}$ |
| medium | $7.135 \times 10^{-3}$ | $4.505 \times 10^{-3}$ | $1.164 \times 10^{-2}$ | $-3.008 \times 10^{-5}$ |
| fine | $7.038 \times 10^{-3}$ | $3.294 \times 10^{-3}$ | $1.033 \times 10^{-2}$ | $2.345 \times 10^{-5}$ |
| ultrafine | $6.947 \times 10^{-3}$ | $2.542 \times 10^{-3}$ | $9.488 \times 10^{-3}$ | $-8.102 \times 10^{-6}$ |
| CFL3D | -- | -- | $8.192 \times 10^{-3}$ | $-6.850 \times 10^{-6}$ |
| Fluent | $6.910 \times 10^{-3}$ | $1.866 \times 10^{-3}$ | $8.774 \times 10^{-3}$ | $2.529 \times 10^{-5}$ |

Table 3: Summary of the global aerodynamic coefficients computed with LBM for different grid resolutions at $\alpha = 10º$. CFL3D values are taken from the NASA turbulence modeling resource (https://turbmodels.larc.nasa.gov/)

| Mesh | $C_{Df}$ | $C_{Dp}$ | $C_D$ | $C_L$ |
| --- | --- | --- | --- | --- |
| medium | $6.781 \times 10^{-3}$ | $3.127 \times 10^{-2}$ | $3.805 \times 10^{-2}$ | 1.0458 |
| fine | $6.746 \times 10^{-3}$ | $2.261 \times 10^{-2}$ | $2.935 \times 10^{-2}$ | 1.0846 |
| ultrafine | $6.757 \times 10^{-3}$ | $1.674 \times 10^{-2}$ | $2.349 \times 10^{-2}$ | 1.1055 |
| CFL3D | $6.204 \times 10^{-3}$ | $6.060 \times 10^{-3}$ | $1.223 \times 10^{-2}$ | 1.0911 |
| Fluent | $6.206 \times 10^{-3}$ | $7.144 \times 10^{-3}$ | $1.335 \times 10^{-2}$ | 1.0973 |

Figure 8 summarizes the grid-convergence study for both angles of attack. A nearly linear trend is observed for the pressure drag component with respect to mesh spacing, whereas the skin friction drag remains nearly constant across the different wall lattice resolutions. The total drag is consisten overpredicted relative to the grid-converged CFL3D reference solution, primarily due to an overestimation of the pressure-drag component. This discrepancy is largely attributable to the limited resolution of the Cartesian grid in the leading edge region, which strongly influences the pressure distribution.

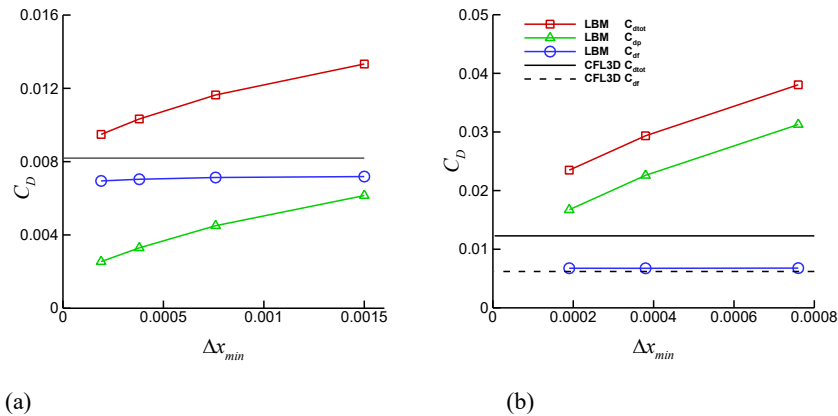

(a)          (b)

Figure 8. Drag convergence study: Total drag is indicated by square symbols, friction drag by circles and pressure drag by triangles. Solid black and dashed lines are the CFL3D total and friction drag, respectively, taken from *[57]*. (a) $\alpha = 0º$. (b) $\alpha = 10º$. Note that the friction drag computed with the LBM is only weakly dependent on the grid resolution.

### 3.1.4 Flow field comparison

LBM computations on the fine grid described in Table 1 have been also compared with RANS solutions obtained using a standard second-order finite-volume solver with the Spalart-Allmaras model. The RANS mesh is a structured-like grid with high resolution inside the boundary layer, generated following common best-practice guidelines. Figure 9(a) and Figure 9(b) provide views of the LBMx and Fluent grids near the trailing-edge region. The computed turbulent viscosity ratio fields are shown for $\alpha = 0º$ in Figure 9(c) and Figure 9(d), and for $\alpha = 10º$ in Figure 9(e) and Figure 9(f). Overall, good agreement in turbulence levels is observed despite differences in grids (structured vs Cartesian) and modeling approaches (RANS vs LBM-RANS). It should be noted that the RANS solutions employ a resolved near-wall treatment, whereas LBM results are obtained using wall functions. For $\alpha = 10º$, the turbulent viscosity near the upper surface close to the trailing edge is slightly underpredicted in the LBM-RANS results, which can be partly attributed to under-resolution in this region in which the boundary layer thickens across coarser grid levels.

Two rakes were defined for velocity probing, at $x/c = 0.85$ for $\alpha = 0º$ and $x/c = 0.5$ for $\alpha = 10º$, as indicated in Figure 9(c-d). Figure 10 presents the normalized velocity profiles and turbulent eddy viscosity ratios extracted along these rakes. For $\alpha = 0º$, the velocity profile at $x/c = 0.85$ (Figure 10(a)) is in very good agreement with the RANS solution, with both boundary-layer thickness and profile shape reasonably well captured. Small discrepancies are observed in the turbulent viscosity ratio (Figure 10(b)), which are expected given the differences in modeling approach, grid resolution, and wall treatment. Note that a spurious peak in the turbulence level appears outside the boundary layer, where the turbulence intensity should be close to the freestream value. This is a numerical artifact that occurs at small regions with different grid resolution levels and is convected downstream, and is due to the Lax-Wendroff discretization, as discussed in Section 2.2. This artifact can be mitigated by employing a first-order upwind discretization for the turbulence model, as discussed in more detail in Section 3.3 or by employing a more sophisticated second order discretization, as described in Section 2.2. Despite these differences, the overall turbulent boundary layer features are well captured by the LBM-RANS approach. Figure 10(c) and Figure 10(d) present the results for $\alpha = 10°$, with the rake positioned at $x/c = 0.5$. This station was selected away from the leading edge to minimize possible non-equilibrium effects associated with the strong pressure gradients in that region. Despite the significant acceleration and deceleration of the flow around the leading edge, the velocity profile shows very good agreement with the RANS solution obtained on a highly resolved grid. The turbulent viscosity ratio is also reasonably well predicted, suggesting that the proposed boundary condition is consistent with both the turbulence model and the wall function formulation.

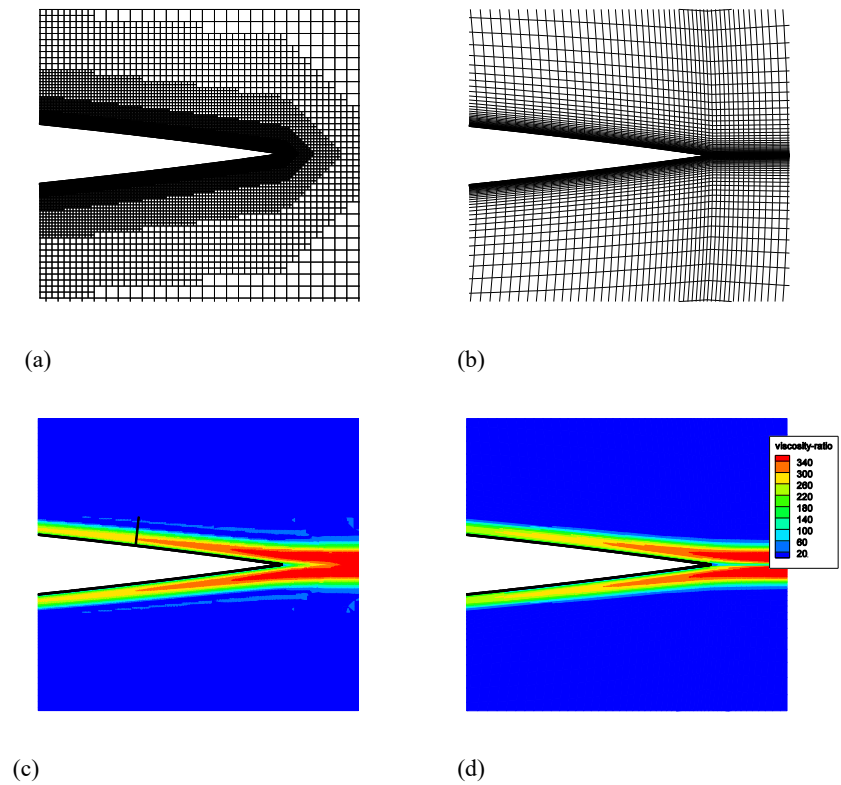

(a)       (b)

(c)       (d)

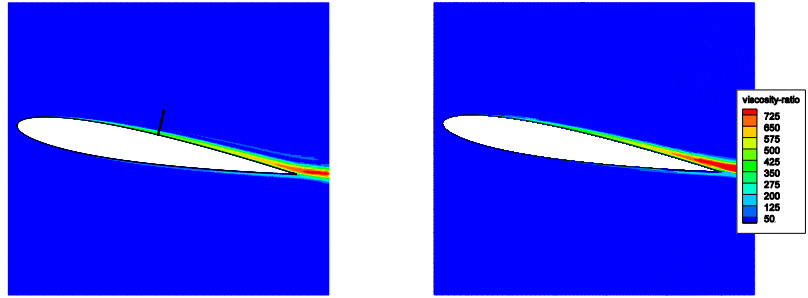

(e)                      (f)

Figure 9. Comparison of turbulent viscosity ratio fields obtained with the LBM VN-slip-velocity bounce-back condition and Fluent. (a) LBM fine mesh for $\alpha = 0º$, (b) Fluent mesh, (c) LBM turbulent viscosity ratio for $\alpha = 0º$, (d) Fluent turbulent viscosity ratio for $\alpha = 0º$, (e) LBM turbulent viscosity ratio for $\alpha = 10º$, and (f) Fluent turbulent viscosity ratio for $\alpha = 10º$. The position of the probing rakes are indicated with black lines in panels (c) and (e).

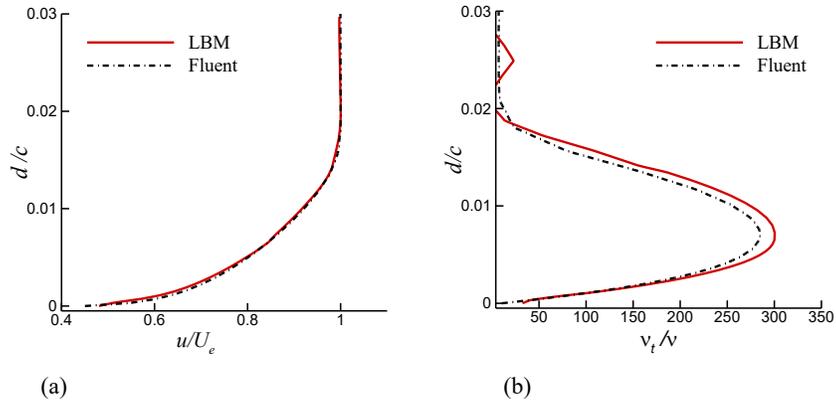

(a)                      (b)

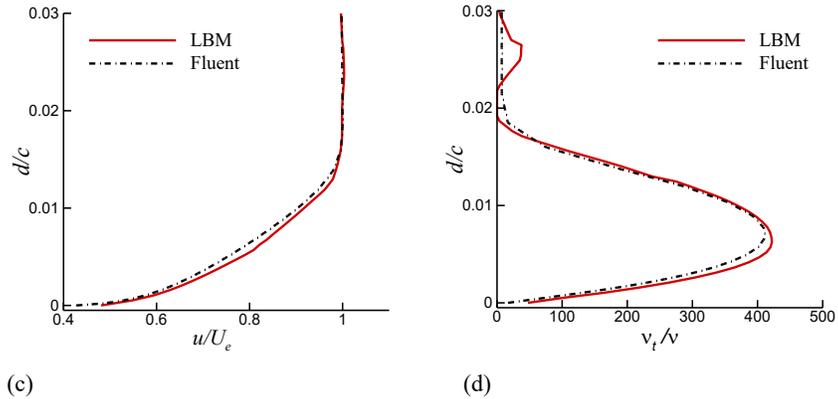

(c)               (d)

Figure 10. Comparison of velocity and turbulent eddy-viscosity ratio as a function of wall-normal distance $d$, obtained with LBM and Fluent RANS solver. (a) velocity profile for $\alpha = 0°$ at $x/c = 0.85$, (b) turbulent viscosity ratio for $\alpha = 0°$ at $x/c = 0.85$, (c) velocity profile for $\alpha = 10°$ at $x/c = 0.5$, (d) turbulent viscosity ratio for $\alpha = 10°$ at $x/c = 0.5$.

### 3.1.5 Comparison of bounce-back based methods: Virtual node *vs* Nishimura's method

In this section, the proposed bounce-back slip-velocity VN method is compared with Nishimura's approach within the LBM-RANS framework. It should be noted that Nishimura et al. [33] do not provide sufficient details to reproduce their method exactly. In particular, no information is given on how the probe distance $d_{WR}$ is determined, other than that it lies within a specified range. However, this range may result in many probe points (R) being located inside a boundary cell, rendering the interpolation of flow variables unreliable. Additionally, the interpolation method used to obtain the flow variables at the probe point is not described. Nishimura et al. performed their simulations with iWMLES, using Spalding's wall function to determine the wall shear and a simple mixing-length model to compute the turbulent viscosity at the reference point. In the present work, their approach has been adapted to the RANS turbulence framework. Therefore, the comparison refers to the behavior of Nishimura's formulation when applied directly within the RANS context.

The key difference with the VN method is that steps 3 and 4 of the procedure described in section 2.3 are omitted, and a linear velocity profile is assumed from the reference point (R) to the wall for extrapolating the slip velocity, as in Nishimura's approach, i.e.

$$u_W = u_R - \frac{u_\tau^2}{\nu + \nu_t\big|_R} d_{WR} \qquad (31)$$

In addition, the reference distance used in (31) is prescribed as $d_{WR} = 1.75$, corresponding to the upper bound of the range indicated by Nishimura and the value adopted by Liang et al. [34], who also implemented Nishimura's method. Figure 11 presents the time history of the friction drag for $\alpha = 0º$ on the NACA0012 using the fine grid. Nishimura's approach predicts a friction drag $Cd_f$ = 0.0083, corresponding to an overestimation of approximately 20% relative to the Fluent's reference value 0.00691. In contrast, The VN-method yields 0.00708, slightly over but considerably closer to Fluent's value.

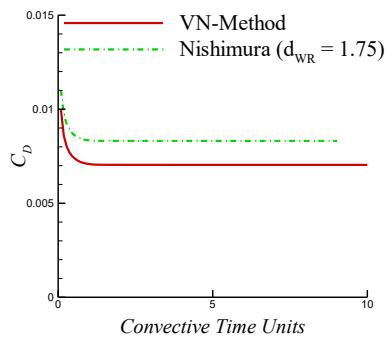

Figure 11. Comparison of the skin friction drag coefficient predicted by Nishimura's bounce-back method and the VN slip velocity bounce-back method for the NACA0012 at $\alpha = 0º$ using the fine mesh resolution.

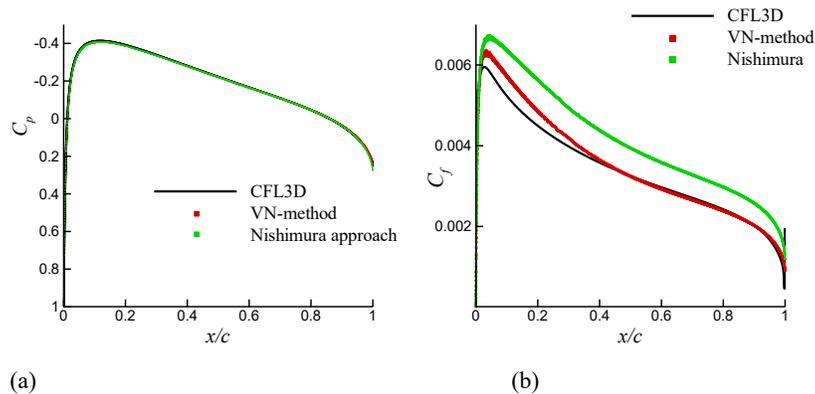

(a)            (b)

Figure 12. Comparison of the surface data obtained with Nishimura's and VN methods. (a) pressure distribution coefficient (b) skin friction coefficient.

Figure 12 presents a comparison of the surface data obtained using the two approaches. The pressure coefficient predicted by Nishimura's method compares well with the reference data (Figure 12(a)), which can be attributed to the minimal influence of the very thin boundary layer on the pressure distribution at this high Reynolds number. In contrast, the differences between Nishimura's approach and the VN method are more pronounced in the skin-friction distribution (Figure 12(b)). Here, the skin-friction coefficient is overpredicted by approximately 28% at $x/c = 0.5$. This discrepancy arises from the inaccurate prediction of the velocity profile, and consequently of the velocity gradients, within the boundary layer, as illustrated in Figure 13(a). The turbulent viscosity levels are also underestimated by roughly 25% compared to Fluent reference values.

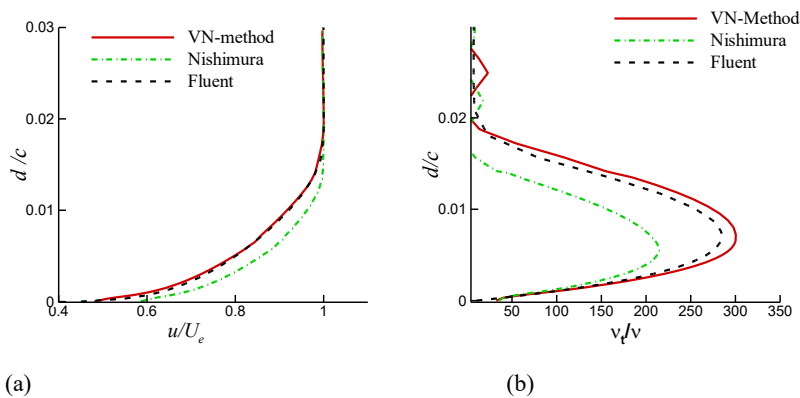

(a)            (b)

Figure 13. Comparison of the boundary layer predictions at $x/c = 0.85$ for the VN an Nishimura wall treatments: (a) streamwise velocity profile and (b) turbulent viscosity ratio.

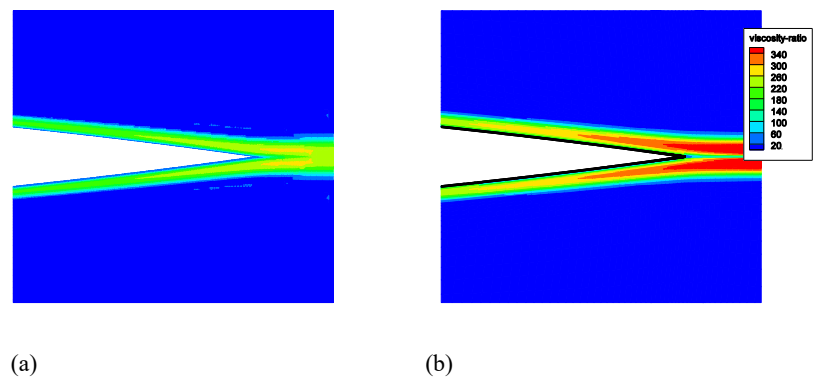

(a)            (b)

32ACCEPTED MANUSCRIPT

Physics of Fluids

This is the author's peer reviewed, accepted manuscript. However, the online version of record will be different from this version once it has been copyedited and typeset.
PLEASE CITE THIS ARTICLE AS DOI: 10.1063/5.0306421

AIP Publishing

Figure 14. Assessment of turbulent viscosity ratio fields obtained with Nishimura's bounce-back condition: (a) Nishimura's method and (b) Fluent RANS solution.

Figure 14 shows contour maps of the turbulent viscosity ratio compared with Fluent reference values. It can be seen that the predicted turbulent boundary-layer thickness is over 25% smaller than those obtained with Fluent and the VN method (see Figure 9 (c-d))

Overall, these results for this simple test case highlight the significant impact of the wall treatment on RANS simulation outcomes. The introduction of the VN method, through a reinterpretation of the bounce-back approach using Chapman-Enskog analysis [21], leads to a substantial improvement in the predicted results. Nishimura's method shows good agreement with experimental pressure distributions for the MD30P30N configuration using iWMLES, consistent with the results shown in Figure 12(a). However, a more detailed evaluation of the turbulence quantities, such as the surface time-averaged skin-friction coefficients and boundary-layer velocity profiles, would be required to fully assess the accuracy of the turbulent wall model in the iWMLES context.

### 3.2 MD30P30N airfoil

LBM computations for flow past the McDonnell Douglas 30P30N three-element airfoil are presented at several flow conditions. LBM results are compared against Fluent RANS computations and experimental data from NASA Low Turbulence Pressure Tunnel (LTPT) [70]. The geometry (which is shown in Figure 20) and experimental data used for these comparisons are available on the AIAA 4th High-Lift Prediction Workshop webpage [71]. This same configuration has previously been used for assessing RANS and LES models for computing high-lift flows [72] [73] [74] [75] [76].

Within LBM-based CFD methods, this configuration has been used in numerous works that can be classified according to the type of boundary treatment used. Using the wet-node approach with non-equilibrium function extrapolation, Maeyama et al. [17] presented iWMLES results for the MD30P30N under JAXA experimental conditions [77], focusing primarily on slat noise prediction. While the acoustic noise predictions show good agreement with the experiments, the computed surface time-averaged pressure distributions on the airfoil exhibit significant discrepancies.

Using an improved wet-node regularized approach, Degrigny [7] reported detailed simulations (including boundary-layer velocity profiles and turbulent viscosity levels) for the MD-30P30N airfoil using ProLB. The study compared hybrid RANS/LES and RANS (Spalart-Allmaras) turbulence models at high Reynolds number ($Re = 9 \times 10^6$) for two angles of attack ($\alpha = 8.10º$ and $\alpha = 16.21º$). The LBM-RANS results correspond to the same angles of attack considered in the present study and are thus directly comparable. Focusing on their LBM-RANS results, overall agreement with the experimental data is obtained. Nevertheless, a significant overprediction of suction pressure is observed on the upper surfaces of all three elements at the lower angle of attack.

Duda et al. [78] presented results with PowerFLOW (using the volumetric-based boundary scheme) employing VLES turbulence modeling in combination with an undisclosed, proprietary method to predict laminar/turbulent transition, although no detailed comparisons with experimental data were provided for pressure, skin friction distributions, or boundary-layer velocity profiles. At lower Reynolds number ($Re = 1.7 \times 10^6$), Satti [79] used PowerFlow in VLES mode to investigate the configuration with the same conditions as in the experiments conducted in the Basic Aerodynamic Research Tunnel (BART) at NASA Langley Research Center including the tunnel walls. They focused primarily on aeroacoustic noise predictions rather than detailed aerodynamic performance, reporting reasonable agreement between the computed flow velocity field in the slat cove region and computed 2D/3D URANS and experimental data. Unfortunately, only mean surface pressure distributions were analyzed in detail and compared against numerical (2D URANS) and experimental data.

With regard to slip-velocity bounce-back-based methods, Ishida [32], Nishimura [33], and Liang [34] presented LBM-WMLES simulations of the turbulent flow around the MD-30P30N at $Re = 1.7 \times 10^6$. As noted in Section 1, their work focused primarily on slat-noise prediction. Ishida and Liang performed 2D simulations, which are not physically realistic for this problem, as the turbulence models they employed are not valid in two dimensions. Nishimura carried out 3D iWMLES simulations for two angles of attack and reported good agreement with experimental pressure distributions. However, as previously mentioned, no details were provided regarding the prediction of turbulent flow quantities.

Coreixas [80] employed a simple staircase bounce-back method without any wall model, imposing only a no-slip velocity at the wall. His simulation of the MD30P30N airfoil at $Re = 1.7 \times 10^6$ and Mach number 0.17 exhibited a strong sensitivity to grid resolution and large discrepancies in the slat-surface pressure distribution. He concluded that an accurate curved-wall treatment and an appropriate turbulent wall model are essential for this configuration.

Finally, an additional point of comparison can be found in the recent work by Husson et al. [23] [26], who carried out a comprehensive study of a similar high lift configuration to the one studied here (LEISA 2) using a wet-node regularized-based approach within the framework of the hybrid RANS/LES ZDES approach.

The aerodynamic flow around high-lift airfoils involves complex physical phenomena that present significant challenges for conventional RANS solvers with near-wall resolved turbulence models [81]. In particular, the use of wall functions is expected to limit the accuracy of predicted flow features, as they inherently simplify the near-wall physics and may struggle to capture detailed effects under conditions characterized by strong pressure gradients, flow separation and laminar-turbulent transition, particularly on the slat surface. Although it is difficult to distinguish modeling errors associated with the RANS and wall function approach from those arising from the LBM boundary-condition formulation, assessing the performance of the proposed wall-boundary scheme in such a complex geometry remains of considerable interest. The intricate geometrical features, such as thin gaps, pronounced curvatures, cove corners, and sharp trailing edges, provide a stringent test of the robustness and fidelity of the boundary condition implementation.

In this work, the following flow conditions are used. The freestream Mach number is 0.2 and the Reynolds number, on the stowed element chord, is 9 million. The angles of attack are $\alpha$ = 8.10º and 16.21º. For higher angles of attack, it has been reported that three-dimensional effects and laminar-to-turbulent boundary-layer transition at the elements should be included to improve predictive accuracy [82].

An automatic Cartesian grid for the MD-30P30N airfoil was generated with the features summarized in Table 4. The surface resolution corresponds to that used for the fine NACA0012 grid described in section 3.1.3. The computational domain extent is [120c, 120c] based on the retracted airfoil chord, with the airfoil located at the center, and tests indicate that further extension has negligible impact on the results. At the trailing edge of the main element, the boundary-layer thickness is 0.015c for $\alpha$ = 8.10º. The estimated number of grid points within the boundary layer is approximately 20, with $y_V$ ≈ 200 at the virtual node. No additional refinement was applied at the leading edges of the airfoil elements, since multi-zone, surface-based geometric refinement is not yet implemented in LBMx. The simulations were conducted using acoustic scaling for the time step, with a fully turbulent model applied throughout the flow field. The LBM near-wall model described in Section 2.3 was employed consistently on all element surfaces, including the slat and main-wing coves. Reference RANS computations were performed using Fluent 2023R1 on a relatively coarse hybrid grid, representative of those typically employed in aerodynamic design phases. The standard Spalart-Allmaras turbulence model was used for these calculations, with grid parameters summarized in Table 4. This comparison strategy allows us to assess the predictive capabilities of the proposed LBMx wall-treatment approach against a widely used industrial baseline, thereby highlighting the potential of adaptive Cartesian LBM methods as a viable alternative for aerodynamic design and analysis.

Table 4: Grid properties for the MD-30P30N test case at $\alpha$ = 16.21º

| Mesh | Δx | No. of elements | Octree levels | No. points in BL at $x/c$ = 0.85 | $y_V^+$ ($x/c$ = 0.85) |
|---|---|---|---|---|---|
| $\alpha$ = 8.1º | 4.6×10$^{-4}$ | 129303 | 14 | ≈ 20 | 205 |
| $\alpha$ =16.2º | 4.6×10$^{-4}$ | 130306 | 14 | ≈ 20 | 196 |
| Fluent (2023-R1) | 0.5×10$^{-6}$ | 60150 | -- | 30+ | <1 |

Figure 15 illustrates the time history of the aerodynamic forces for the $\alpha$ = 8.10º test case. The flow reaches a steady, RANS-like convergence state after approximately 90-100 time convective units. The time history for the $\alpha$ = 16.21º case (not shown) exhibits a similar behaviour. Figure 16 presents the normalized velocity magnitude and turbulent eddy-viscosity ratio fields, respectively, for the $\alpha$ = 16.21º demonstrating smooth transitions across the different mesh levels in both macroscopic velocity and turbulence fields.

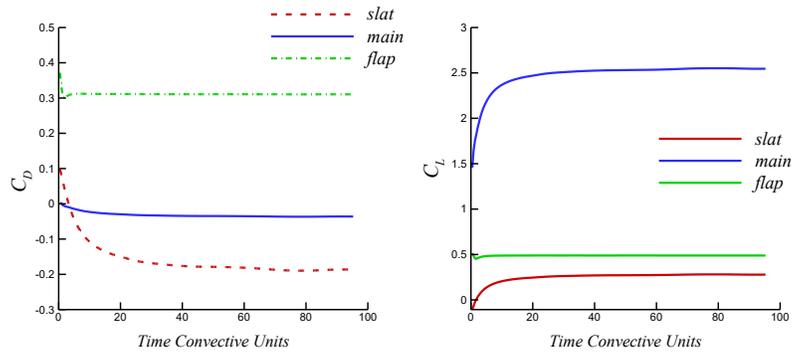

Figure 15. Time history of total lift and drag coefficients for the three high lift airfoil elements at $\alpha = 8.10º$

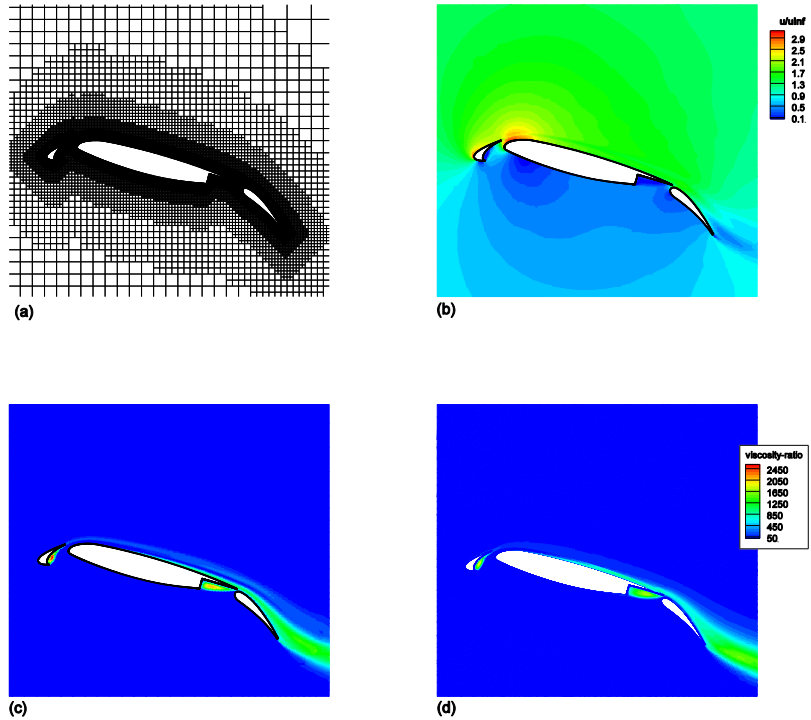

Figure 16. LBM computation of the MD-30P30N airfoils at $\alpha$ = 16.21º. (a) Adaptive Cartesian grid, (b) velocity magnitude normalized by the reference far-field velocity, (c) LBM turbulent eddy-viscosity ratio, and (d) Fluent turbulent eddy-viscosity ratio.

### 3.2.2 Surface data results

Computed surface pressure and skin-friction coefficient distributions from LBM-RANS and conventional finite volume RANS are shown in Figure 17 and Figure 18 for $\alpha$ = 8.10º and $\alpha$ = 16.21º, respectively. Experimental results from NASA LTPT [71] are also included for comparison. Overall, the pressure coefficient distributions exhibit good agreement with the experimental measurements. An exception is observed for the slat at $\alpha$ = 8.10º (Figure 17(a)), where the pressure suction level is clearly overpredicted relative to the experiments. This discrepancy has also been reported by several authors [72] [73] [75] [76], and its origin remains uncertain, likely resulting from a combination of multiple factors as discussed in [72]. Interestingly, the Fluent RANS results predicts the slat suction level more accurately; however, this outcome should be interpreted with caution, as it differs from most RANS results reported in the literature. In contrast, the agreement between the LBM and Fluent RANS pressure distributions is excellent for both the main wing and flap elements as can be seen in Figure 17(c) and 17(e).

A key challenge in verifying the proposed boundary scheme lies in the accurate prediction of the skin-friction coefficient across the three airfoil elements. Figure 17(b), 17(d) and 17(f) present the computed skin-friction distributions from LBM and conventional RANS for the slat, main wing, and flap, respectively. LTPT measurements are available only at two points on the main element and at a few locations on the flap. The LBM predictions with the proposed boundary scheme yield smooth skin-friction distributions across all three elements, showing overall good agreement with the wall-resolved RANS results obtained using the SA turbulence model. Small discrepancies appear on the slat and flap, primarily due to under-resolved boundary layers at the leading edges of these components. Furthermore, while the RANS simulation captures a small separation at the flap trailing edge, this feature is absent in the LBM predictions, likely due to the limitations of the equilibrium wall model employed. On the other hand, the agreement over the main element is excellent, particularly downstream of the leading edge where the boundary layer recovers and approaches equilibrium conditions, similar to observations from the NACA0012 airfoil test case.

For $\alpha$ = 16.21º, the results follow a similar trend. Despite the strong pressure gradients at the slat leading edge, the pressure coefficient distribution predicted by the LBM solver shows very good agreement with both RANS and experimental data, as illustrated in Figure 18(a-c-e). The largest discrepancy arises in the skin-friction distribution over the slat (Figure 18(b)), which can again be attributed to the limitations of the equilibrium wall model. In particular, the strong favorable/adverse pressure gradients near the slat leading edge affect the accuracy of the friction velocity prediction in this region. Further downstream, closer to the slat trailing edge, the LBM-predicted skin friction converges toward the RANS solution as the pressure gradient relaxes. Overall, the LBM approach reproduces the surface aerodynamic quantities with good accuracy and smoothness, demonstrating the robustness of the proposed boundary treatment in conjunction with the immersed Cartesian grid methodology.

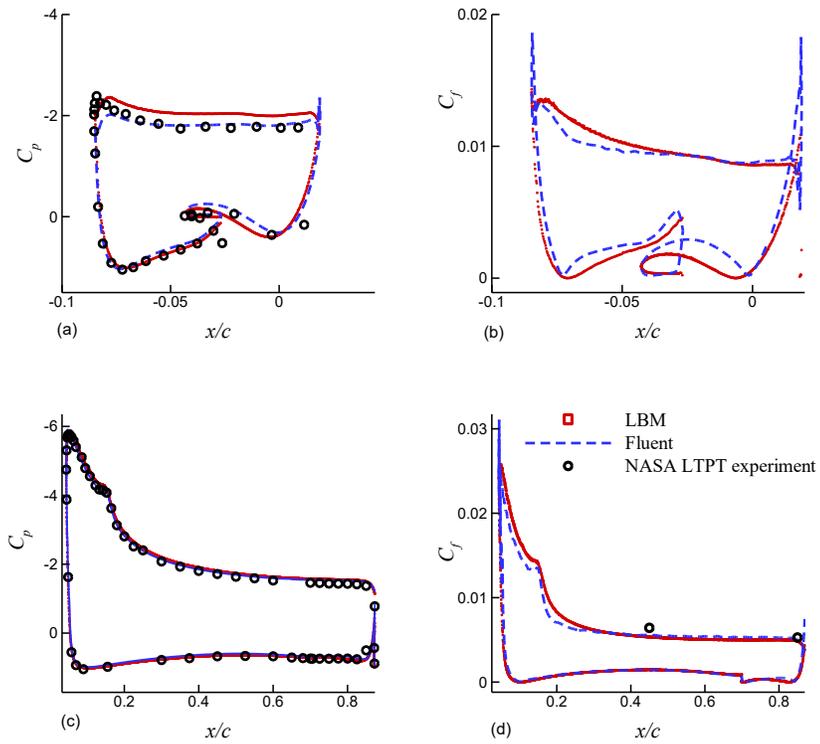

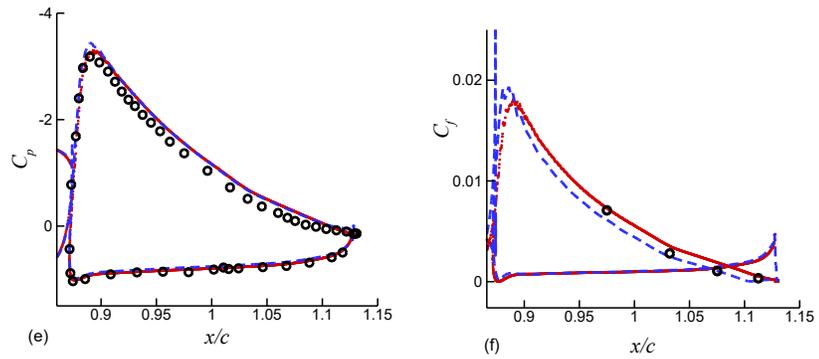

Figure 17. LBM and RANS computed surface pressure and skin-friction coefficients compared with experimental data from LTPT [70] for $\alpha = 8.10º$. (a-c-e) Pressure coefficient for slat, main and flap elements (b-d-f) skin-friction coefficient, skin-friction coefficient. Experimental results are indicated by circular symbols.

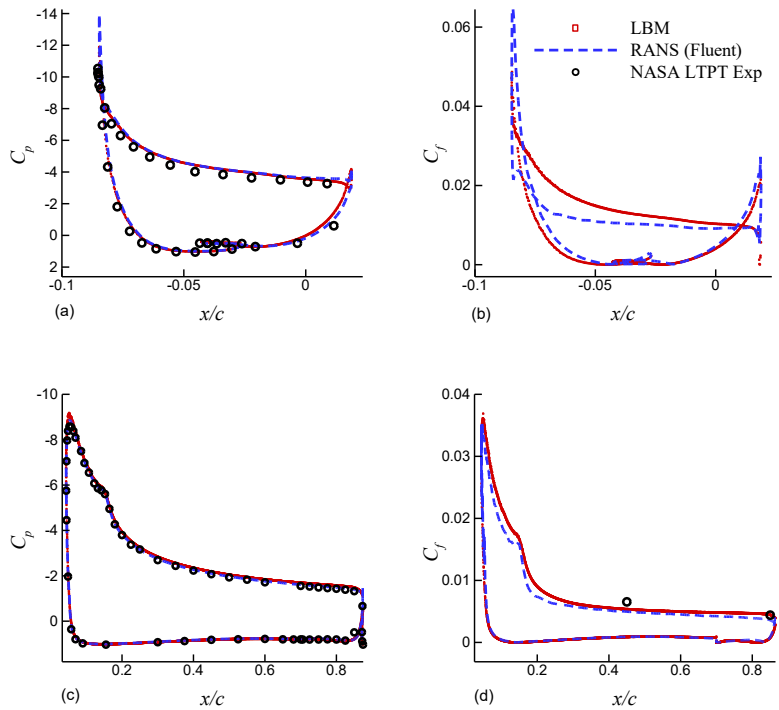

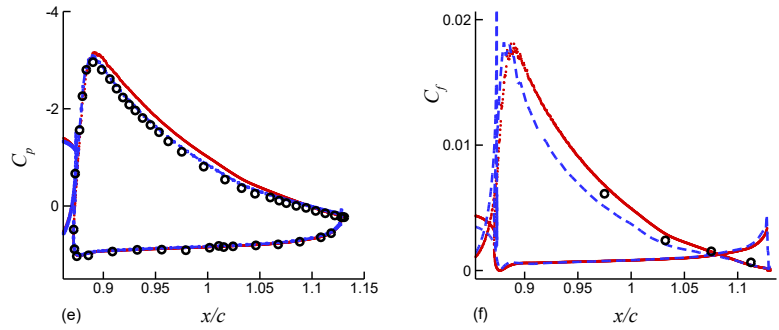

Figure 18. LBM and RANS computed surface pressure and skin-friction coefficients compared with experimental data from LTPT [70] for $\alpha$ = 16.21º. (a-c-e) Pressure coefficient for slat, main and flap elements (b-d-f) skin-friction coefficient for slat, main and flap elements respectively. Experimental results are indicated by circular symbols.

### 3.2.2 Grid sensitivity study for the MD-30P30N configuration

A grid sensitivity analysis, analogous to the one performed for the NACA0012 test case, has been conducted for the MD-30P30N configuration at $\alpha$ = 8.1º to evaluate the robustness and accuracy of the proposed VN method in a more physically and geometrically complex flow. Medium, fine and ultrafine grid resolutions have been computed, using three same values of $\Delta x_{min} / c$ as those used in the NACA0012 test case.

Table 5 summarizes the obtained aerodynamic force coefficients and compares them with the reference Fluent simulation. The trends observed in the NACA0012 case are also seen here. First, the friction drag coefficient shows very weak sensitivity to grid refinement: the relative difference between the ultrafine mesh and the Fluent prediction is about 4%, whereas for the medium-resolution LBM grid the discrepancy is about 9%. In contrast, the pressure drag systematically decreases with grid refinement because the pressure-suction peaks near the leading edges are more accurately captured on finer meshes. For the medium grid, the pressure drag and total drag differ from the Fluent values by only 0.7% and 1.9%, respectively. However, the total drag remains significantly overpredicted when compared with both the experimental data and the FUN3D-FV results reported by NASA [76], obtained using the SA-neg turbulence model on an extremely fine 2D mesh (609,000 elements). This discrepancy arises because the total drag in this configuration is dominated by pressure drag, which depends on a delicate balance between the positive contribution from the flap element and the negative contributions from the main element and the slat (see Figure 15). Consequently, even small over- or under-predictions in any of these components can lead to relatively large errors in the total drag. In contrast, the lift coefficient is consistently well predicted across all grids, showing a relative deviation of approximately 2.2% from the NASA reference value obtained with the FUN3D-FV solver reported in [76] for the ultrafine grid resolution.

Table 5: MD-30P30N force results for the grid study performed at $\alpha$ = 8.1º. The last column provides the corresponding computed $y_v^+$. Experimental value is taken from AIAA High-Lift prediction workshop IV (https://aiaa-hlpw.org/HLPW/index-workshop4.html)

| Mesh | Ncells | $C_{Df}$ | $C_{Dp}$ | $C_D$ | $C_L$ | $y_v^+$ ($x/c$ = 0.85) |
|---|---|---|---|---|---|---|
| medium | 65409 | 8.61×10$^{-3}$ | 9.90×10$^{-2}$ | 1.09×10$^{-1}$ | 3.198 | 406 |
| fine | 129303 | 9.99×10$^{-3}$ | 7.75×10$^{-2}$ | 8.75×10$^{-2}$ | 3.313 | 206 |
| ultrafine | 266896 | 9.99×10$^{-3}$ | 6.79×10$^{-2}$ | 7.79×10$^{-2}$ | 3.412 | 105 |
| Fluent | 60150 | 9.55×10$^{-3}$ | 9.97×10$^{-2}$ | 1.07x10$^{-1}$ | 3.264 | < 1 |
| Experiment | | -- | -- | -- | 3.135 | -- |

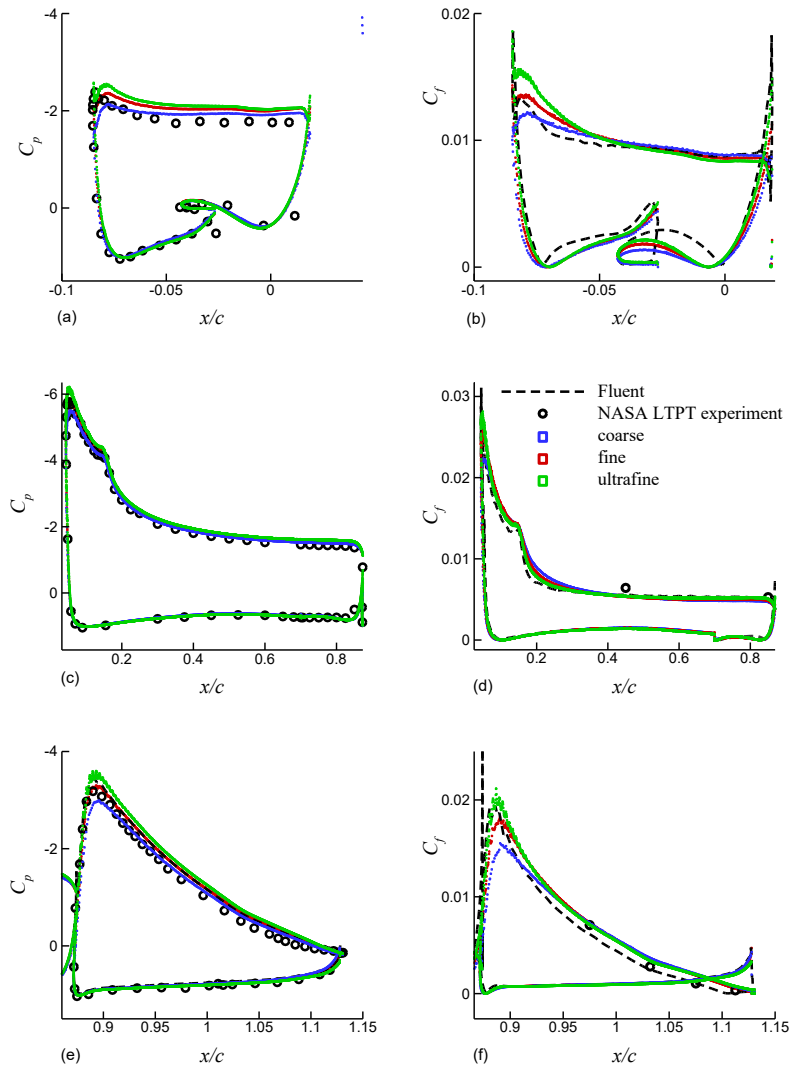

Figure 19. Grid convergence study for the MD30P30N test case at $\alpha = 8.1°$. (a-c-e): pressure coefficient for slat, main and flap elements, respectively. (b-d-f): skin-friction coefficient for slat, main and flap elements, respectively. Experimental results are represented with black circles.

### 3.2.3 Flow field comparison

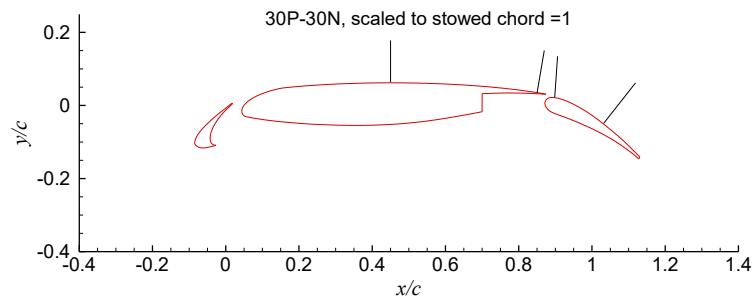

Figure 20. Geometry and locations of the stations used for comparison of velocity profiles with experimental measurements.

To further evaluate the accuracy of the proposed LBM wall treatment, turbulent boundary-layer velocity profiles were compared with experimental measurements from [71]. Figure 20 shows the selected stations: two on the main element ($x/c$ = 0.45 and $x/c$ = 0.85) and two on the flap ($x/c$ = 0.889 and $x/c$ = 1.032).

Figure 21 presents the comparisons for the main-element stations at both computed angles of attack. Overall, the agreement between LBM-RANS and experiments is very good, particularly at the trailing-edge station ($x/c$ = 0.85), where upstream effects that are not explicitly resolved in the simulation (e.g., laminar-turbulent transition, local under-resolution) have largely dissipated. At the mid-chord location ($x/c$ = 0.45), the LBM-RANS results slightly overpredict the boundary-layer thickness (Figure 21(a) and Figure 21(c)), consistent with previously reported findings [81] and likely attributable to the fully turbulent assumption used in this study. At $\alpha$ = 8.10º (Figure 21(b)), the trailing-edge velocity profile ($x/c$ = 0.85) is well captured, while the slat-wake signature is clearly predicted at both stations on the main element. This feature is absent in the experimental data, but has also been observed in RANS simulations with the Spalart-Allmaras model using conventional finite-volume solvers [76] [81]. Other investigations have shown that high-fidelity turbulence models provide a more accurate description of the slat-wake velocity deficit for this configuration [7] [74] [76]. At $\alpha$ =16.21º (Figure 21(c) and 21(d)), the slat wake is present in the experimental results but its velocity deficit is overpredicted by both LBM-RANS and RANS simulations. This overprediction, also reported in prior studies, is likely linked to the absence of laminar-turbulent transition modeling on the slat surface, which has a significant impact on downstream wake spreading.

At the flap stations (Figure 22), the boundary layer at $x/c$ = 0.889 is too thin to be resolved accurately by the wall function. Further downstream ($x/c$ = 1.032), the turbulent boundary-layer thickness is reasonably predicted for both angles of attack. The main-element wake is underresolved due to limited grid resolution in the O-type Cartesian mesh (Figure 22 (a)), leading to slight diffusion of shear layers across grid levels. Localized refinement would likely improve wake prediction, though this capability is not currently implemented.

Despite these limitations, the VN slip-velocity bounce-back scheme combined with the turbulence wall model successfully captures the dominant turbulent flow features over the main element and flap. The results demonstrate the robustness and reliability of the proposed method for complex, multi-element high-lift airfoils.

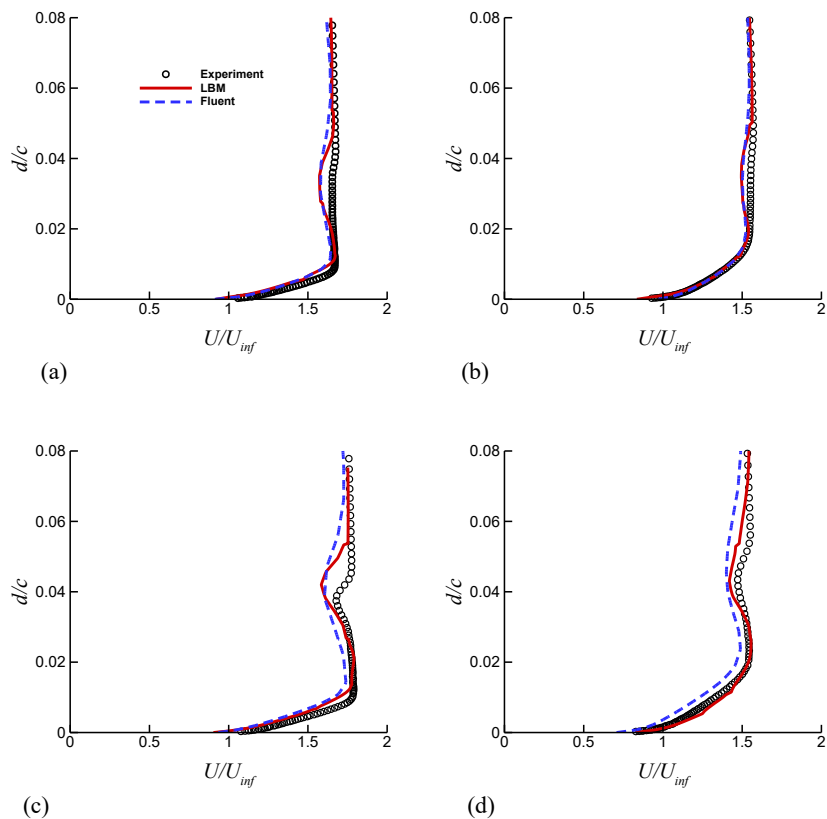

Figure 21. Comparison of LBM velocity profiles, Fluent RANS computations and experimental data for two different angles of attack and four stations located on the main airfoil element. (a) $\alpha = 8.10°$, $x/c = 0.45$; (b) $\alpha = 8.10°$, $x/c = 0.85$; (c) $\alpha = 16.21°$, $x/c = 0.45$; (d) $\alpha = 16.21°$, $x/c = 0.85$

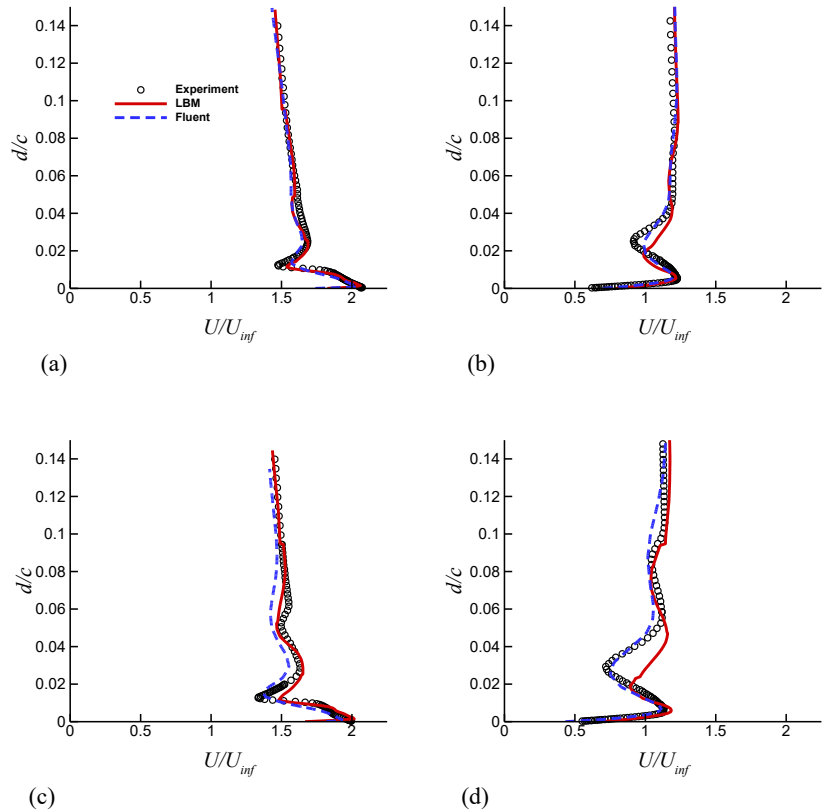

Figure 22. Comparison of LBM velocity profiles, Fluent RANS computations and experimental data for two different angles of attack and at four stations located on the flap element. (a) $\alpha$ = 8.10º, $x/c$ = 0.889; (b) $\alpha$ = 8.10º, $x/c$ = 1.03; (c) $\alpha$ = 16.21º, $x/c$ = 0.889; (d) $\alpha$ = 16.21º, $x/c$ = 1.03

### 3.3 Influence of the turbulence model discretization of convective terms

In this section, we examine the influence of the finite-difference discretization order used for the convective terms of the SA turbulence model. The LW scheme may introduce numerical oscillations in the early stages of the computation, which typically decay as the solution converges. This behavior is observed for the NACA0012 airfoil test. For this particular case, a persistent spurious oscillation develops outside the boundary layer near the trailing edge, in the vicinity of the interface between different grid-resolution levels, as previously discussed in Section 2.2 and 3.1.4. This is illustrated in Figure 23(a). This numerical artifact is eliminated when a first-order upwind scheme is used for the convective terms of the turbulence model as seen in Figure 23(b). However, the first-order scheme also introduces additional numerical dissipation, leading to lower predicted turbulent viscosity ratios compared with the LW discretization. In addition, the turbulent boundary-layer edge becomes noticeably more diffused when the upwind discretization is employed (Figure 23(b)). Despite these differences in the turbulence quantities, the impact on the mean velocity profile is minimal, as illustrated in Figure 24(a). Nonetheless, the widening of the turbulent boundary-layer edge with the first-order scheme remains evident in Figure 24(b).

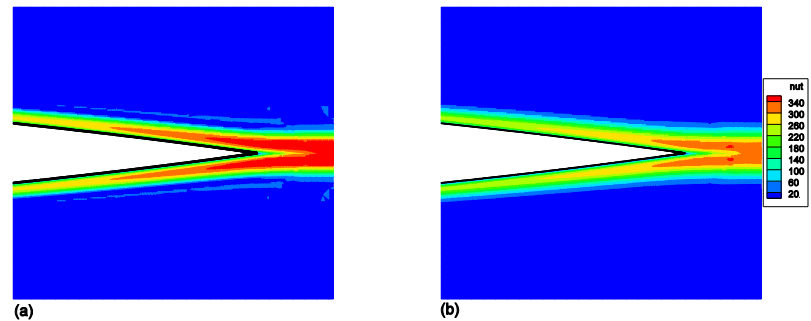

Figure 23. Comparison of turbulent viscosity ratio fields for the NACA0012 airfoil at $\alpha = 0º$ obtained with different discretizations of the convective terms of the SA model (a) Lax-Wendroff second order discretization. (b) 1st order upwind discretization.

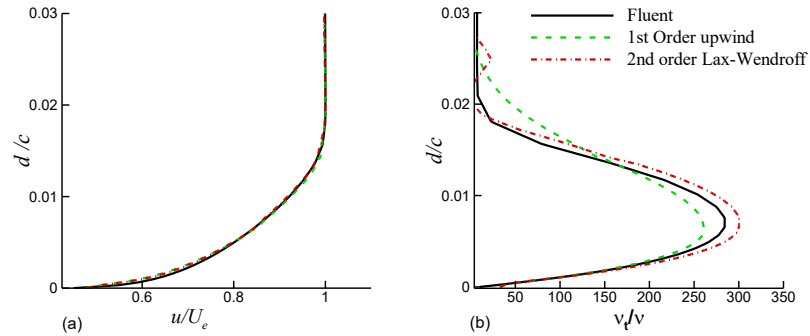

Figure 24. Comparison of velocity (a) and turbulent viscosity ratio (b) profiles at $x/c = 0.85$ obtained with different discretizations of the convective term of the SA model for the NACA0012 airfoil at $\alpha = 0º$. Fluent results with second-order upwind discretization are also shown for reference.

In general, a first-order upwind discretization is sufficient for accurately simulating attached boundary layers, consistent with the behavior commonly observed in classical finite-volume RANS solvers. However, our results indicate that, as in conventional CFD practice, employing a second-order discretization becomes advisable for more complex configurations where strong shear layers develop outside the attached boundary-layer regions, as occurs in the MD-30P30N test case. In this configuration, the slat wake interacts with both the main-element wake and with the boundary layer on the upper surface of the flap. Under such conditions, first-order upwind predictions may lack accuracy unless the grid is already sufficiently refined in those regions or additional adaptive refinement is applied in a second step to properly resolve the shear layers. This behavior is illustrated in Figure 25, which shows the turbulent viscosity ratio fields for the MD-30P30N configuration computed on the fine grid at $\alpha = 8.1º$ using the two discretization schemes for the turbulence model. The LW discretization yields turbulence levels comparable to the second-order Fluent solution, while producing less diffusive shear-layer interfaces, Overall, the flow and turbulence fields produced by the second-order LW discretization resemble the second-order Fluent solution more closely. Finally, in contrast to the NACA0012 case, no spurious oscillations were observed for this configuration when using the LW discretization.

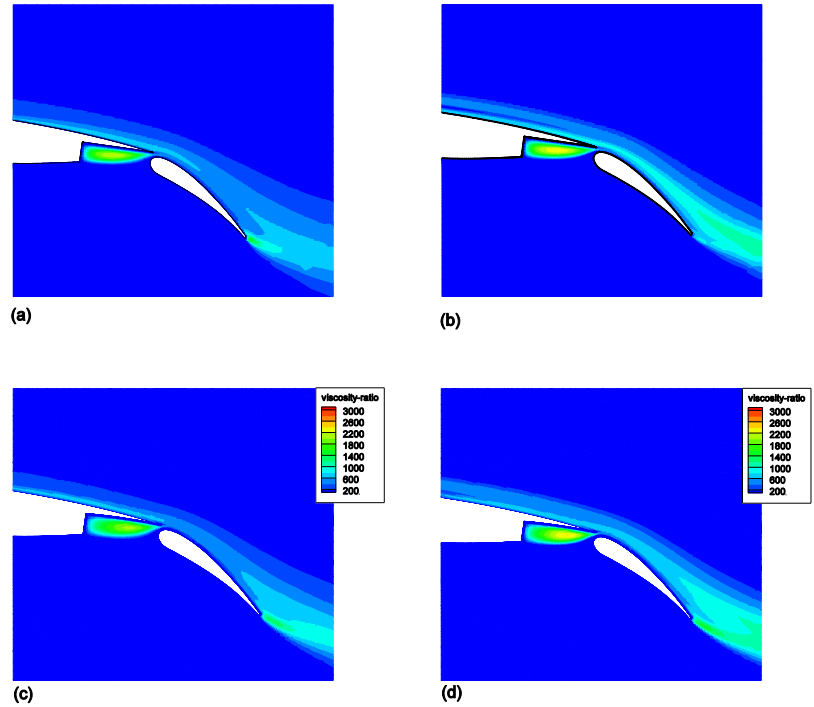

Figure 25. Comparison of turbulent viscosity ratio contours obtained with different convective-term discretizations employed in the turbulence model for the MD30P30N configuration at $\alpha = 8.1°$ (a) LBM with first-order upwind discretization. (b) LBM with second-order Lax-Wendroff discretization. (c) Fluent with first-order upwind discretization. (d) Fluent with second-order upwind discretization.

## 4. Conclusions

We have presented a novel VN slip-velocity bounce-back immersed wall boundary treatment within the LBM-RANS framework for high-Reynolds-number turbulent flows on multilevel adaptive Cartesian meshes. Building on previous work for flat surfaces [21], which derives the slip velocity from an analysis of how the bounce-back condition, augmented with a wall momentum correction term, enforces the macroscopic flow boundary condition at the boundary node, the method has been extended to curved geometries. The approach couples turbulence modeling, wall functions, and LBM boundary conditions in a manner designed to suppress oscillations in key surface quantities.

The method has been validated on two configurations: a NACA0012 airfoil and a multielement MD30P30N airfoil. For the NACA0012 case, LBM-RANS results show excellent agreement with standard CFD solvers (CFL3D and Fluent) in terms of surface quantities, turbulent boundary-layer profiles and eddy-viscosity distributions. A grid-convergence study indicates robust behavior, with skin-friction differences of the order of 3% between the coarsest and finest grids. For the MD30P30N case, the scheme demonstrates robustness for turbulent flows over complex geometries, providing reliable predictions of both surface and flow-field quantities despite the limitations of the wall model (equilibrium wall function). An additional grid-convergence study confirms the low grid sensitivity of the predicted skin-friction drag. In both test cases, the approach yields smooth surface pressure and skin friction coefficients.

The proposed method requires minimal auxiliary geometric data, making it potentially well-suited for GPU-based implementations, and naturally enforces the boundary condition via link-wise bounce-back without additional ad-hoc treatments. Future work will be dedicated to include non-equilibrium effects in the wall model and extending the approach to three-dimensional high-fidelity turbulence simulations.

**Funding**: This research was funded by INTA under the grant IDATEC (IGB21001)

**Acknowledgment**: We thank Dr. Jiménez Varona and Mr. Olalla Sanchez for providing the Fluent simulations of the NACA0012 and MD30P30N test cases